\documentclass[12pt,preprint]{aastex62}
\shorttitle{Nature vs.  Nurture}
\shortauthors{Safsten \& Dawson}

\usepackage{mathtools}
\usepackage{lineno}
\usepackage{amsmath}
\usepackage{etoolbox}
\usepackage{appendix}
\usepackage{xcolor}
\usepackage{hhline}
\newcommand*\linenomathpatch[1]{%
  \cspreto{#1}{\linenomath}%
  \cspreto{#1*}{\linenomath}%
  \csappto{end#1}{\endlinenomath}%
  \csappto{end#1*}{\endlinenomath}%
}

\linenomathpatch{equation}
\linenomathpatch{gather}
\linenomathpatch{multline}
\linenomathpatch{align}
\linenomathpatch{alignat}
\linenomathpatch{flalign}

\begin{document}

\title{Nature vs. Nurture: Investigating the Effects of Measurement Uncertainties in the Assessment of Potential Trends Between Planetary and Stellar Properties}
\received{}
\revised{}
\correspondingauthor{Emily D. Safsten}
\email{eds36@psu.edu}
\author[0000-0002-3425-7803]{Emily D. Safsten}
\affiliation{Department of Astronomy \& Astrophysics, Center for Exoplanets and Habitable Worlds, The Pennsylvania State University, University Park, PA 16802, USA}
\author[0000-0001-9677-1296]{Rebekah I. Dawson}
\affiliation{Department of Astronomy \& Astrophysics, Center for Exoplanets and Habitable Worlds, The Pennsylvania State University, University Park, PA 16802, USA}

\begin{abstract}
Correlations between planetary and stellar properties, particularly age, can provide insight on planetary formation and evolution processes.  However, the underlying source of such trends can be unclear, and measurement uncertainties and small sample sizes can leave doubt as to whether an observed trend truly exists.  We use a Bayesian framework to examine how uncertainties in measured parameters influence the odds ratios of competing hypotheses for the source of an observed trend.  We analyze three reported trends from the literature.  In each application, while uncertainties do affect the numerical value of the odds ratios, our conclusions remain the same whether or not uncertainties are taken into account: hot Jupiter eccentricities are circularized over time, obliquities of hot Jupiter hosts are driven by stellar temperature, and there is not enough evidence to favor a trend of 2:1 orbital resonances with age over a chance relation.  Updated samples for the 2:1 resonances and obliquities cases do not change the original conclusions.  Simulated 2:1 resonance data show that sample size may be more important than measurement precision for drawing a firm conclusion.  However, if 2:1 resonances get disrupted on a wide range of timescales, an age trend will be inherently difficult to confirm over a chance relation, even with a large sample.  For some applications, full incorporation of measurement uncertainties may be too computationally expensive, making it preferable to use the framework without uncertainties and perform additional tests to examine the effects of highly uncertain measurements.
\end{abstract}

\section{Introduction}
A good grasp of the properties of stars is essential for understanding the planets they may host, both because stars and planets form contemporaneously from the same original cloud of gas and dust, and because most observations of exoplanets are done indirectly through study of their host stars.  Correlations between observed stellar and planetary properties can provide clues about planetary formation and evolution \citep{christiansen2019}.  In particular, a trend in a planetary property due to stellar age indicates that the planetary property evolves over time, while correlations driven by other stellar parameters (such as effective temperature or mass) may indicate that formation conditions are important for shaping the observed distribution of the planetary property.  However, the stellar properties themselves are often interrelated (for example, stars change temperature as they age), which can make it difficult to know the true underlying source of an observed trend.  Additionally, proposed trends between planetary and stellar properties may suffer from small sample sizes or large measurement uncertainties, which can leave doubt about whether an apparent trend is real or not.

In \cite{safstenetal2020}, hereafter Paper 1, we developed a Bayesian framework to assess the strengths of trends of planetary orbital properties with stellar ages.  This framework is designed to compare the evidence for two competing hypotheses using an odds ratio, to determine which hypothesis the data better supports.  In particular, we defined three hypotheses for the source of an observed trend between a planetary property and a stellar property: that the planetary property is due to the system age and thus evolves over time (Nurture); that the planetary property is due to an observed system parameter other than age (Nature); and that the planetary property is independent of both age and other observed system parameters (Chance).  We then used this framework to investigate proposed trends of three planetary properties from the literature: the disruption of 2:1 orbital resonances \citep{kz2011}, the obliquities of stars with hot Jupiters (e.g., \citealt{winn2010,triaud2011}), and the eccentricities of hot Jupiters (e.g., \citealt{quinn2014}).  We found that obliquities of hot Jupiter hosts are most likely driven by stellar temperature, and that eccentricities of hot Jupiters are driven by age (i.e., they evolve over time due to tidal circularization).  Our result for the 2:1 orbital resonances case was inconclusive.

The Bayesian framework in Paper 1 lacked a formal incorporation of measurement uncertainties.  For each case that we examined in Paper 1, we performed additional tests to examine the potential effects of highly uncertain measurements, and in all cases found our overall conclusions to be the same.  However, we wish to fully incorporate uncertainties in order to make our framework more general and accurate.  In particular, stellar ages typically have large uncertainties (see \citealt{soderblom2010} for a review), though recent observations by the Gaia spacecraft \citep{gaiamission} have enabled better constraints on many stellar parameters, including ages (e.g. \citealt{berger2018,berger2020}).  In this work, we investigate the effects of measurement uncertainties on the odds ratios we obtain.

Additionally, in both the 2:1 resonances and stellar obliquities cases, we analyzed the original data sets used by \cite{kz2011} and \cite{triaud2011} (the studies that originally reported the age trends), respectively, in order to directly compare our results with those of the original papers.  More exoplanets have been discovered and characterized since these data sets were compiled, and we wish to apply our framework to updated samples to see how the odds ratios may change, particularly in the 2:1 resonances case, for which our original analysis was inconclusive.  In addition to having a greater number of planets, an updated sample can allow us to take advantage of studies that have homogeneously derived ages for planet host stars.

Our aim in this paper is twofold.  First, we further generalize the Bayesian framework described in Paper 1 by formally incorporating measurement uncertainties (Section \ref{section:generalframework}).  We then apply this updated framework to the same datasets used in Paper 1 and compare the results to our original odds ratio calculations.  Second, as the original samples for the 2:1 resonances and obliquities cases were compiled a decade ago, we apply our framework to the most recent available data on 2:1 resonances and stellar obliquities.  The analysis for the case of 2:1 resonances is in Section \ref{section:resonances}, for the case of stellar obliquities is in Section \ref{section:obliquities}, and for the case of hot Jupiter eccentricities is in Section \ref{section:eccentricities}.  We then conclude in Section \ref{section:conclusion}.

\section{General Framework}
\label{section:generalframework}
The Bayesian framework developed in Paper 1 is designed to compare the strengths of competing hypotheses given the available data.  The hypotheses represent possible explanations for an apparent correlation between a certain planetary property, denoted $X_{\rm p}$, and other observed system properties.  To assess the relative strengths of two hypotheses $H_{\rm A}$ and $H_{\rm B}$, we need to compute an odds ratio (Eqn. 1 in Paper 1):
\begin{equation}
\label{eqn:oddsratio}
    \frac{p(H_{\rm A}|\{X_{\rm p,i},t_{\rm {\star},i},{\bf X}_{\rm{ob},i}\})}{p(H_{\rm B}|\{X_{\rm p,i},t_{\rm {\star},i},{\bf X}_{\rm{ob},i}\})}=\frac{\int[\prod\limits_{\rm i}{p(X_{\rm p,i},t_{\rm {\star},i},{\bf X}_{\rm{ob},i}|{\bf Y},H_{\rm A})]p({\bf Y})d{\bf Y}}}{\int[\prod\limits_{\rm i}{p(X_{\rm p,i},t_{\rm {\star},i},{\bf X}_{\rm{ob},i}|{\bf Y},H_{\rm B})]p({\bf Y})d{\bf Y}}}~ \frac{p(H_{\rm A})}{p(H_{\rm B})},
\end{equation}
the ratio of the probabilities of each hypothesis given the available data, where $t_{\star}$ is the stellar age and ${\bf X}_{\rm ob}$ is observed system parameters other than age and $X_{\rm p}$.  This requires us to compute, for each individual system and under each hypothesis, the likelihood $p(X_{\rm p},t_{\star},{\bf X}_{\rm{ob}}|{\bf Y})$ that the system exists with a certain set of $X_{\rm p}$, age, and other system parameters, given the hyperparameters (population-wide variables) ${\bf Y}$.  It also requires us to specify our a priori beliefs on the relative strengths of $H_{\rm A}$ and $H_{\rm B}$, represented by $p(H_{\rm A})/p(H_{\rm B})$. Hereafter, we give equal weight to each hypothesis, so this term cancels out.  Note that we also drop the $i$ subscript, with the understanding that variables other than hyperparameters refer to individual system parameters.

In Paper 1, we defined three hypotheses for the source of the observed distribution of $X_{\rm p}$.  First, the Nurture hypothesis says that the planetary property $X_{\rm p}$ is driven by the age of the system, i.e. it evolves over time.  Second, the Nature hypothesis says that $X_{\rm p}$ is due to a system parameter other than age, such as stellar temperature, and there is no time evolution of $X_{\rm p}$.  Finally, the Chance hypothesis says $X_{\rm p}$ is independent of both age and other system parameters.  For a more detailed description of the framework, including a more thorough explanation of the Nurture, Nature, and Chance hypotheses, we refer the reader to Section 2 in Paper 1.

We first give the basic equation for the likelihood from which the rest of the framework was derived (Eqn. 2 from Paper 1):
\begin{equation}
    \label{eqn:oldgeneral}
    p(X_{\rm p},t_{\star},{\bf X}_{\rm ob}|{\bf Y})=\int p(X_{\rm p},t_{\star},{\bf X}_{\rm ob},{\bf X}_{\rm nob}|{\bf Y}) d{\bf X}_{\rm nob},
\end{equation}
where ${\bf X}_{\rm nob}$ contains other unobserved system parameters that may be relevant for the question at hand.  This equation implicitly assumes that all observed parameters are measured at their true values with zero uncertainty.  For each of the observed parameters, we need to include a term in Eqn. \ref{eqn:oldgeneral} for the probability of measuring the observed value of the parameter given the true value.  We then marginalize over the true parameter values.  Let $X_{\rm p,obs}$, $t_{\rm \star,obs}$, and ${\bf X}_{\rm ob,obs}$ represent the observed values of the planetary property of interest, the stellar age, and the other observed system properties, respectively, for an individual planetary system, and let $X_{\rm p}$, $t_{\star}$, and ${\bf X}_{\rm ob}$ represent the true values of those quantities, respectively.  Then, with uncertainties included, Eqn. \ref{eqn:oldgeneral} becomes:
\begin{align}
    \label{eqn:generaluncertainties}
    p(X_{\rm p,obs},t_{\rm \star,obs},{\bf X}_{\rm ob,obs}|{\bf Y})&=\iiiint p(X_{\rm p,obs},t_{\rm \star,obs},{\bf X}_{\rm ob,obs}|X_{\rm p},t_{\star},{\bf X}_{\rm ob})\nonumber \\ & \quad \times p(X_{\rm p},t_{\star},{\bf X}_{\rm ob},{\bf X}_{\rm nob}|{\bf Y}) dX_{\rm p}dt_{\star}d{\bf X}_{\rm ob}d{\bf X}_{\rm nob}.
\end{align}
The $p(X_{\rm p},t_{\star},{\bf X}_{\rm ob},{\bf X}_{\rm nob}|{\bf Y})$ term can then be modified for each hypothesis as described in Paper 1.  This yields the following equations for each hypothesis.

The equations for the Nurture hypothesis introduce two unobserved parameters, $X_{\rm p0}$ and $t_{\rm e}$.  $X_{\rm p0}$ is the initial value of $X_{\rm p}$, which is important because $X_{\rm p}$ evolves over time in this hypothesis.  The variable $t_{\rm e}$ is the evolutionary timescale, representing the time it takes for a planetary system to go from $X_{\rm p0}$ to $X_{\rm p}$.  Then the equation for the Nurture hypothesis, in which $X_{\rm p}$ is driven by the stellar age, $t_{\star}$, is (modified from Eqn. 6 in Paper 1):
\begin{align}
\begin{split}
    p(X_{\rm p,obs},t_{\rm \star,obs},{\bf X}_{\rm ob,obs}|{\bf Y}) &= \iiiint p(X_{\rm p,obs},t_{\rm \star,obs},{\bf X}_{\rm ob,obs}|X_{\rm p},t_{\star},{\bf X}_{\rm ob})\\ &\times\bigg[\iint p(X_{\rm p}|t_\star,t_{\rm e},X_{\rm p0},{\bf Y}) p(t_{\rm e}|{\bf X}_{\rm ob},{\bf X}_{\rm nob},X_{\rm p0},{\bf Y}) \\
    & \qquad\quad \times p(t_\star,{\bf X}_{\rm ob},{\bf X}_{\rm nob}|{\bf Y})p(X_{\rm p0}|{\bf Y})dt_{\rm e} dX_{\rm p0} \bigg] dX_{\rm p}dt_{\star}d{\bf X}_{\rm ob}d{\bf X}_{\rm nob}.
\end{split}
\end{align}

In some applications, $X_{\rm p}$ may take on only two discrete values, which may be represented as $X_{\rm p}=0$ and $X_{\rm p}=1$. The Nurture equation for such a binary $X_{\rm p}$ is (modified from Eqn. 7 in Paper 1):
\begin{align}
\begin{split}
    p(X_{\rm p,obs},t_{\rm \star,obs},{\bf X}_{\rm ob,obs}|f_0,{\bf Y}) &= \iiint \sum_{X_{\rm p}=0}^1 p(X_{\rm p,obs},t_{\rm \star,obs},{\bf X}_{\rm ob,obs}|X_{\rm p},t_{\star},{\bf X}_{\rm ob})\\ &\times\bigg[\int\sum_{X_{\rm p0}=0}^1 p(X_{\rm p}|t_\star,t_{\rm e},X_{\rm p0},{\bf Y}) p(t_{\rm e}|{\bf X}_{\rm ob},{\bf X}_{\rm nob},X_{\rm p0},{\bf Y}) \\
    & \qquad\quad \times p(t_\star,{\bf X}_{\rm ob},{\bf X}_{\rm nob}|{\bf Y})p(X_{\rm p0}|f_0)dt_{\rm e}\bigg] dt_{\star}d{\bf X}_{\rm ob}d{\bf X}_{\rm nob}.
\end{split}
\end{align}
Here, $f_0$ is a hyperparameter representing the population-wide fraction of systems with a particular value of $X_{\rm p0}$, for example, the fraction of systems that start out with a 2:1 orbital resonance.

The Nurture equation for single-direction evolution of binary $X_{\rm p}$, i.e. where $X_{\rm p}$ can evolve from $X_{\rm p0}=1$ to $X_{\rm p}=0$ but not the other way, is (modified from Eqn. 8 in Paper 1):
\begin{align}
    \label{eqn:gennurturebinary}
    p(X_{\rm p,obs},t_{\rm \star,obs},{\bf X}_{\rm ob,obs}|f_0,{\bf Y})=
    \begin{cases}
    \iiint p(X_{\rm p,obs},t_{\rm \star,obs},{\bf X}_{\rm ob,obs}|X_{\rm p}=0,t_{\star},{\bf X}_{\rm ob}) \\ \quad\times \left[1-f_0+f_0\int_0^{t_{\star}}p(t_{\rm e}|X_{\rm p0},{\bf X}_{\rm ob},{\bf X}_{\rm nob},{\bf Y})dt_{\rm e}\right] \\ \quad\times p(t_{{\star}},{\bf X}_{\rm ob},{\bf X}_{\rm nob}|{\bf Y}) dt_{\star}d{\bf X}_{\rm ob}d{\bf X}_{\rm nob} &,X_{\rm p}=0\\
    \iiint p(X_{\rm p,obs},t_{\rm \star,obs},{\bf X}_{\rm ob,obs}|X_{\rm p}=1,t_{\star},{\bf X}_{\rm ob}) \\ \quad\times \left[f_0\int_{t_{\star}}^\infty p(t_{\rm e}|X_{\rm p0},{\bf X}_{\rm ob},{\bf X}_{\rm nob},{\bf Y})dt_{\rm e}\right]\\ \quad\times p(t_{{\star}},{\bf X}_{\rm ob},{\bf X}_{\rm nob}|{\bf Y}) dt_{\star}d{\bf X}_{\rm ob}d{\bf X}_{\rm nob}&,X_{\rm p}=1.
    \end{cases}
\end{align}
In this special case, $f_0$ specifically represents the fraction of systems with $X_{\rm p0}=1$.

The equation for the Nature hypothesis, in which $X_{\rm p}$ is driven by a system parameter other than age, contained in ${\bf X}_{\rm ob}$, is (modified from Eqn. 9 in Paper 1):
\begin{align}
\begin{split}
    \label{eqn:gennature}
    p(X_{\rm p,obs},t_{\rm \star,obs},{\bf X}_{\rm ob,obs}|{\bf Y}) &= \iiiint p(X_{\rm p,obs},t_{\rm \star,obs},{\bf X}_{\rm ob,obs}|X_{\rm p},t_{\star},{\bf X}_{\rm ob})\\ &\times p(X_{\rm p}|{\bf X}_{\rm ob},{\bf Y})p(t_{{\star}},{\bf X}_{\rm ob},{\bf X}_{\rm nob}|{\bf Y}) dX_{\rm p}dt_{\star}d{\bf X}_{\rm ob}d{\bf X}_{\rm nob}.
\end{split}
\end{align}

The equation for the Chance hypothesis, in which $X_{\rm p}$ is not related to either age or other system parameters, and is only determined by random chance, is (modified from Eqn. 10 in Paper 1):
\begin{align}
\begin{split}
    \label{eqn:genchance}
    p(X_{\rm p,obs},t_{\rm \star,obs},{\bf X}_{\rm ob,obs}|{\bf Y}) &= \iiiint p(X_{\rm p,obs},t_{\rm \star,obs},{\bf X}_{\rm ob,obs}|X_{\rm p},t_{\star},{\bf X}_{\rm ob})\\ &\times p(X_{\rm p}|{\bf Y})p(t_{{\star}},{\bf X}_{\rm ob},{\bf X}_{\rm nob}|{\bf Y}) dX_{\rm p}dt_{\star}d{\bf X}_{\rm ob}d{\bf X}_{\rm nob}.
\end{split}
\end{align}

The Chance equation for binary $X_{\rm p}$, i.e. where $X_{\rm p}$ can only take on two discrete values, $X_{\rm p}=0$ or $X_{\rm p}=1$, is (modified from Eqn. 11 in Paper 1):
\begin{align}
    \label{eqn:genchancebinary}
    p(X_{\rm p,obs},t_{\rm \star,obs},{\bf X}_{\rm ob,obs}|f,{\bf Y})=
    \begin{cases}
    \iiint p(X_{\rm p,obs},t_{\rm \star,obs},{\bf X}_{\rm ob,obs}|X_{\rm p}=0,t_{\star},{\bf X}_{\rm ob}) \\ \times (1-f) p(t_{{\star}},{\bf X}_{\rm ob},{\bf X}_{\rm nob}|{\bf Y})dt_{\star}d{\bf X}_{\rm ob}d{\bf X}_{\rm nob} &,X_{\rm p}=0\\
    \iiint p(X_{\rm p,obs},t_{\rm \star,obs},{\bf X}_{\rm ob,obs}|X_{\rm p}=1,t_{\star},{\bf X}_{\rm ob}) \\ \times f p(t_{{\star}},{\bf X}_{\rm ob},{\bf X}_{\rm nob}|{\bf Y})dt_{\star}d{\bf X}_{\rm ob}d{\bf X}_{\rm nob} &,X_{\rm p}=1.
    \end{cases}
\end{align}
In this hypothesis, $f$ is the population-wide fraction of systems with $X_{\rm p}=1$.

In many instances, the uncertainty distributions associated with the measured quantities are assumed to be Gaussian, with reported error bars representing $1\sigma$. Under this assumption, Eqn. \ref{eqn:generaluncertainties} becomes
\begin{align}
    p(X_{\rm p,obs},t_{\rm \star,obs},{\bf X}_{\rm ob,obs}|{\bf Y})&=\iiiint {\rm N}(X_{\rm p},\sigma_{\rm Xp}){\rm N}(t_{\rm \star},\sigma_{\rm t\star}){\rm N}({\bf X}_{\rm ob},\sigma_{\rm Xob})\nonumber\\ &\times p(X_{\rm p},t_{\star},{\bf X}_{\rm ob},{\bf X}_{\rm nob}|{\bf Y}) dX_{\rm p}dt_{\star}d{\bf X}_{\rm ob}d{\bf X}_{\rm nob},
\end{align}
where $\sigma_{\rm Xp}$, $\sigma_{\rm t\star}$, and $\sigma_{\rm Xob}$ are the $1\sigma$ measurement uncertainties on $X_{\rm p,obs}$, $t_{\rm \star,obs}$, and ${\bf X}_{\rm ob,obs}$, respectively.

\begin{figure}[ht]
\centering
    \includegraphics[width=6.5 in]{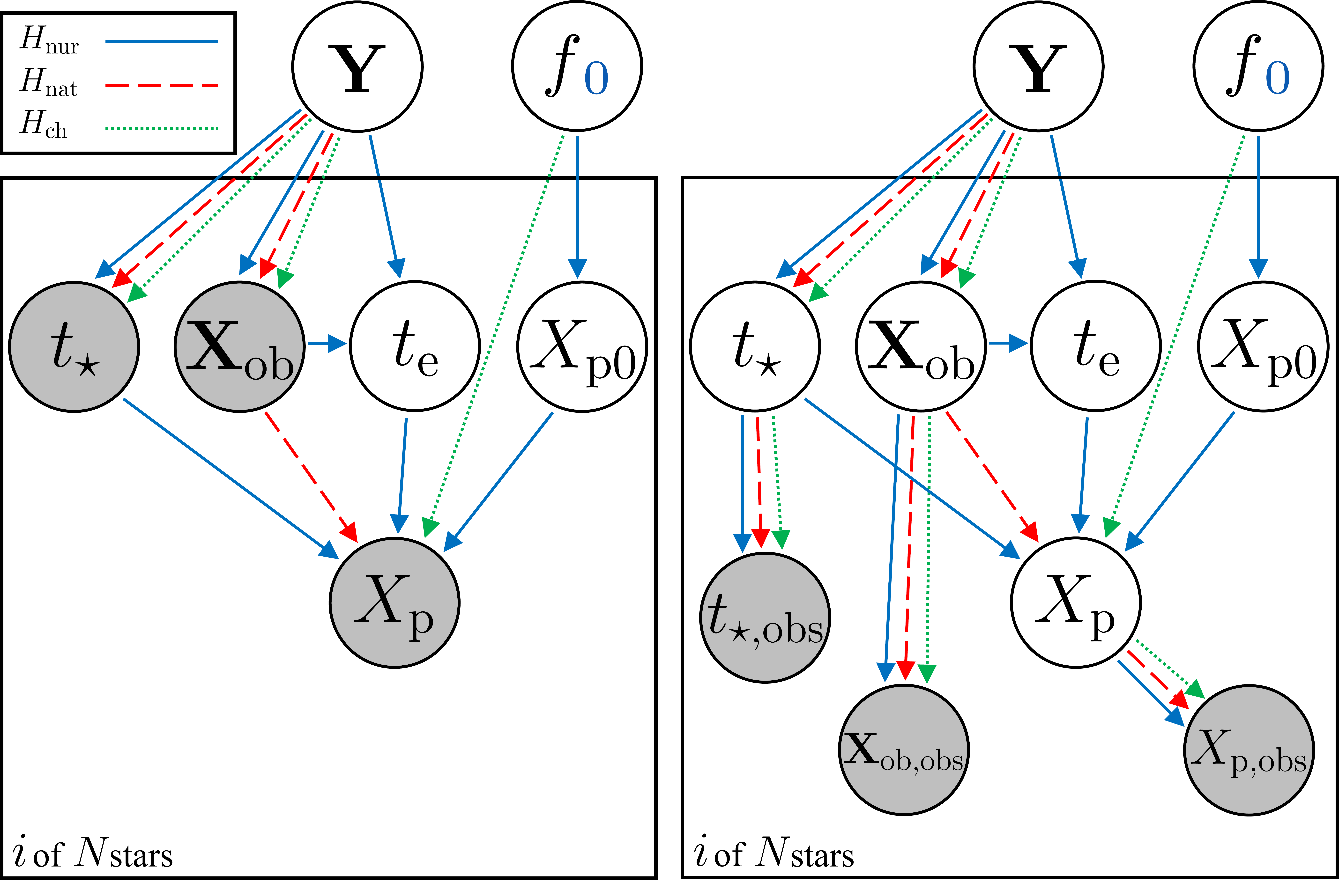}
    \caption{Graphical representation of relationships between parameters under the three hypotheses without uncertainties incorporated (left) and with uncertainties incorporated (right).  For simplicity, we do not show every possible dependence a model may have, but instead tend to focus on those that come up most often in our applications (specifically, we do not explicitly include relevant non-observed quantities ${\bf X}_{\rm nob}$).  Parameter relations under the Nurture, Nature, and Chance hypotheses are shown with blue solid, red dashed, and green dotted lines, respectively.  The ``$i$ of $N$ stars" means that the parameters in the box, or plate, are iterated over each of the $N$ systems in the sample.  Gray circles represent observed parameters:  the observed planetary property of interest ($X_{\rm p}$ on the left, $X_{\rm p,obs}$ on the right), the observed stellar age ($t_{\star}$ on the left, $t_{\rm\star,obs}$ on the right), and other observed parameters (${\bf X}_{\rm ob}$ on the left, ${\bf X}_{\rm ob,obs}$ on the right).  White circles are unobserved individual parameters (on the plate) -- the initial value of $X_{\rm p}$ ($X_{\rm p0}$), the evolutionary timescale ($t_{\rm e}$), and, on the right, the true value of the planetary property of interest ($X_{\rm p}$), the true stellar age ($t_{\star}$), and the true values of other observed parameters (${\bf X}_{\rm ob}$) -- and hyperparameters -- the fraction of systems with a given value of $X_{\rm p}$ ($f$), the fraction of systems with a given value of $X_{\rm p0}$ ($f_0$), and other hyperparameters (${\bf Y}$).  The hyperparameters $f$ and $f_0$ are in the same circle because mathematically, they behave the same way in each hypothesis; they are distinguished by the blue subscript, because $f_0$ occurs in the Nurture hypothesis while the Chance hypothesis uses $f$.}
    \label{fig:generalmodel}
\end{figure}

In Figure \ref{fig:generalmodel}, we display a graphical model showing relationships between parameters and hyperparameters under the three hypotheses without incorporating uncertainties (left) and when uncertainties are incorporated (right).  For simplicity, we do not include every possible parameter dependence, but instead tend to focus on those that most often appear in our applications.

As in Paper 1, we interpret the odds ratios using a scale similar to that of \cite{jeffreys1961} and \cite{kass1995} and consider ratios of $\sim1-10$ to be inconclusive, $\sim10-100$ to be moderately supportive but not decisive, and $\gtrsim100$ to be strong.

\section{2:1 Orbital Resonances}
\label{section:resonances}
Various mechanisms can result in planets locked in orbital resonances.  However, it is not known if these configurations are generally stable over long timescales and how common these configurations are shortly after formation.  Studies such as \cite{thommes2008} and \cite{izid17} have demonstrated resonance disruption through dynamical interactions.  \cite{kz2011} found evidence that systems with 2:1 period commensurabilities were younger than those without. This observed trend suggests that 2:1 resonances may get systematically disrupted over time.  If true, this disruption would have important implications for understanding the stability of planetary systems.  However, follow-up work by \cite{dd2016} explored the initial resonant fraction and disruption timescale required to account for such a trend.  They found that in order to fit the data, all systems would need to start out in resonance.  They would also all need to be disrupted on a timescale similar to the age difference between the 2:1 resonant and nonresonant systems.  Such fine-tuning of the parameters cast doubt as to whether the observed age trend is real or not.

In Paper 1, we applied our Bayesian framework to the \cite{kz2011} data, but our analysis did not find sufficient evidence that 2:1 orbital resonances are systematically disrupted over time.  Instead, we found an odds ratio of the Nurture hypothesis to the Chance hypothesis of 2.2, which is not strong enough to favor either hypothesis.  Here, we revisit the question with our modified framework applied to the original data used by \cite{kz2011} as well as to several recently obtained samples of planets.  We compare evidence for two hypotheses: the Nurture hypothesis, which says that 2:1 resonances are disrupted over time; and the Chance hypothesis, which says that there is no connection between 2:1 resonances and any other system parameters that we consider.  As in Paper 1, we do not test the Nature hypothesis because, to our knowledge, there has not been any proposed correlation between 2:1 resonances and system parameters other than age.  

We treat this case as having a binary $X_{\rm p}$ -- either a system is near a 2:1 resonance, or it is not.  We denote this resonance state as $R$, and we say a system has $R=1$ if it is near a 2:1 resonance and $R=0$ if it is not.  $R$ represents the true 2:1 resonance state of a system, and $R_{\rm obs}$ represents its observed 2:1 resonance state.  The term $p(R_{\rm obs}|R)$ accounts for uncertainties in the measured period ratio and represents the probability of observing a system in a certain resonance state given its true resonance state.  In the samples we analyze, all but a few of the systems are either near a 2:1 resonance or not near a 2:1 resonance at the 3$\sigma$ level.  Accordingly, we will assume $R_{\rm obs}=R$ and drop the $p(R_{\rm obs}|R)$ term in the following equations.  In Appendix \ref{sec:appendix}, we fully account for uncertainties in period ratio, but find our results to be nearly identical to what we obtain here.  We give the stellar age $t_\star$ a uniform prior from 0 to 13.7 Gyr, corresponding to the assumption of a constant star formation rate throughout the age of the universe, and we assume that uncertainties in $t_\star$ are Gaussian.  The general equations applied to this case are as follows.

For the Nurture (time-evolution) hypothesis, we assume that $R$ can only evolve in one direction, from $R=1$ to $R=0$, i.e. resonances may be broken, but not reformed later on.  We also introduce a 2:1 resonance disruption timescale $t_{\rm d}$, which is the time it takes an individual system to go from $R=1$ to $R=0$, and which we give a lognormal prior with mean $\mu$ and standard deviation $\sigma$.  We assign $p(\mu)=U(6,16)$ and $p(\sigma)=U(0,20)$, both in log$_{10}$[yr] space.  We also include $f_0$, the fraction of systems that form near a 2:1 resonance, which we give a uniform hyperprior from 0 to 1.  Note that $f_0$, $\mu$, and $\sigma$ are all hyperparameters.  Then the Nurture equation becomes (derived from Eqn. \ref{eqn:gennurturebinary} and analogous to Eqn. 21 in Paper 1):
\begin{align}
    \label{eqn:resnur}
    p(R_{\rm obs},t_{\rm \star,obs}|\mu,\sigma,f_0)=
    \begin{cases}
    \int dt_\star p(t_{\rm \star,obs}|t_\star) \left[(1-f_0)+f_0\int_0^{t_{\star}} dt_{\rm d} p(t_{\rm d}|\mu,\sigma)\right] p(t_{\star}) &,R_{\rm obs}=0\\
    \int dt_\star p(t_{\rm \star,obs}|t_\star) f_0\left[\int_{t_{\star}}^\infty dt_{\rm d} p(t_{\rm d}|\mu,\sigma)\right] p(t_{\star}) &,R_{\rm obs}=1.
    \end{cases}
\end{align}

In the Chance hypothesis, we have the hyperparameter $f$, which is the overall fraction of systems near a 2:1 resonance.  We give $f$ a uniform hyperprior from 0 to 1.  Then the Chance equation becomes (derived from Eqn. \ref{eqn:genchancebinary} and analogous to Eqn. 23 in Paper 1):
\begin{align}
    \label{eqn:resch}
    p(R_{\rm obs},t_{\rm \star,obs}|f)=
    \begin{cases}
    \int p(t_{\rm \star,obs}|t_{\star})p(t_{\star})(1-f) dt_\star &, R_{\rm obs}=0\\
    \int p(t_{\rm \star,obs}|t_{\star})p(t_{\star}) f dt_\star &, R_{\rm obs}=1.
    \end{cases}
\end{align}

We consider a pair of planets to be near a 2:1 resonance if they have a normalized commensurability proximity (NCP) score of $\delta<0.1$, according to \cite{kz2011}.  The NCP score is defined by \cite{kz2011} as
\begin{equation}
    \delta=2\frac{|r-r_{\rm c}|}{r+r_{\rm c}},
    \label{eqn:ncp}
\end{equation}
where $r$ is the measured period ratio and $r_{\rm c}$ is the period commensurability ratio of interest (2:1, in this case).  If a pair of planets in a given system meets this criterion, the entire system is flagged as being near a 2:1 resonance.  We note that this threshold for a system being near a 2:1 resonance is somewhat arbitrary.  Furthermore, period commensurability alone is not sufficient for a pair of planets to be in resonance -- the resonance angle must also librate.  However, this condition is much more difficult to ascertain than a simple period commensurability.

\subsection{Results}
\label{subsec:resresults}
For this case, we perform integrations in Python using {\tt scipy.integrate.nquad} with default settings.  This integrator uses techniques from the Fortran library QUADPACK; it uses a Clenshaw-Curtis method using Chebyshev moments for finite integration limits, and a Fourier integral if there is an infinite limit.

\begin{figure}[ht]
    \includegraphics[width=3.5 in]{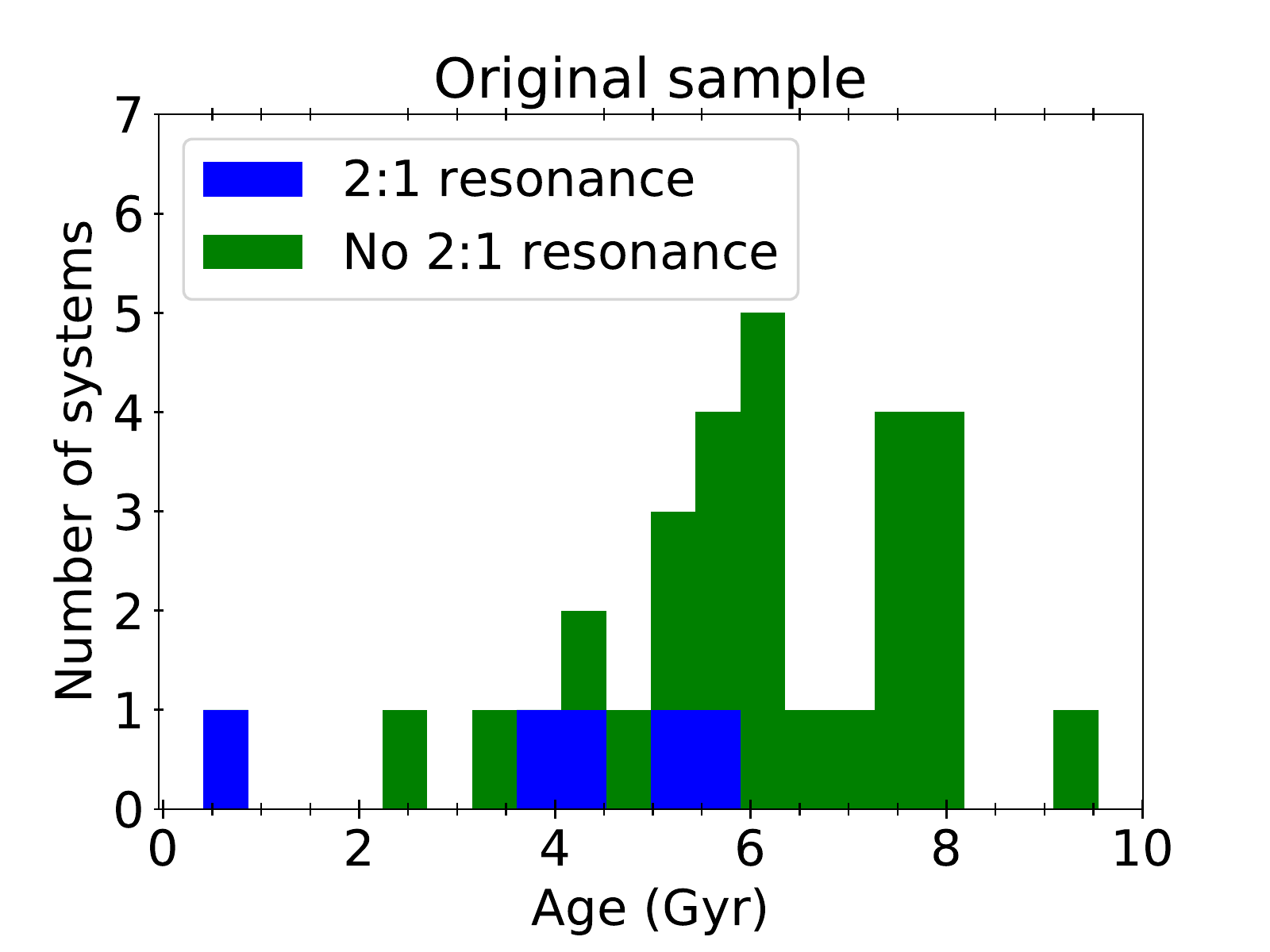}
    \includegraphics[width=3.5 in]{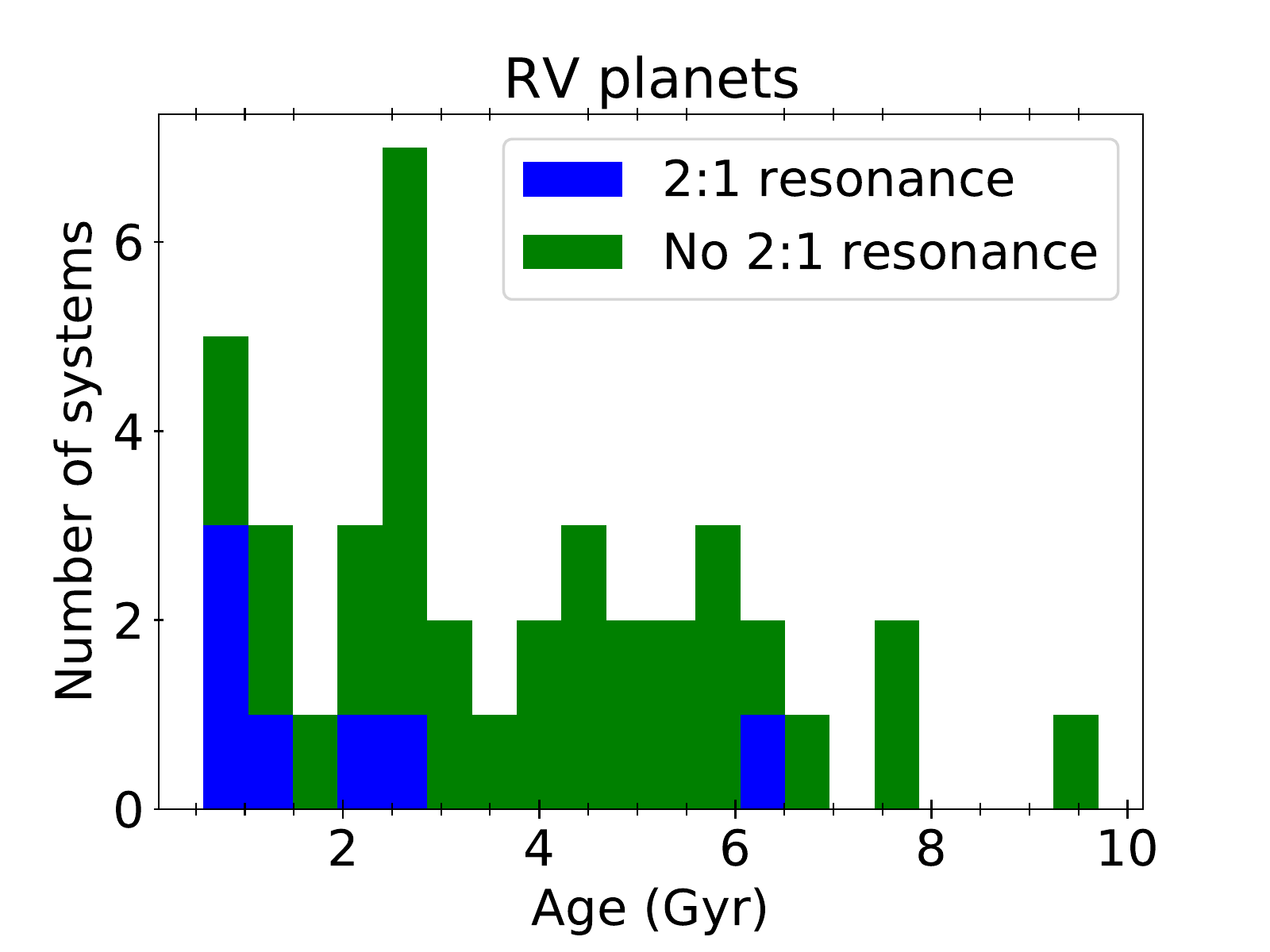}
    \includegraphics[width=3.5 in]{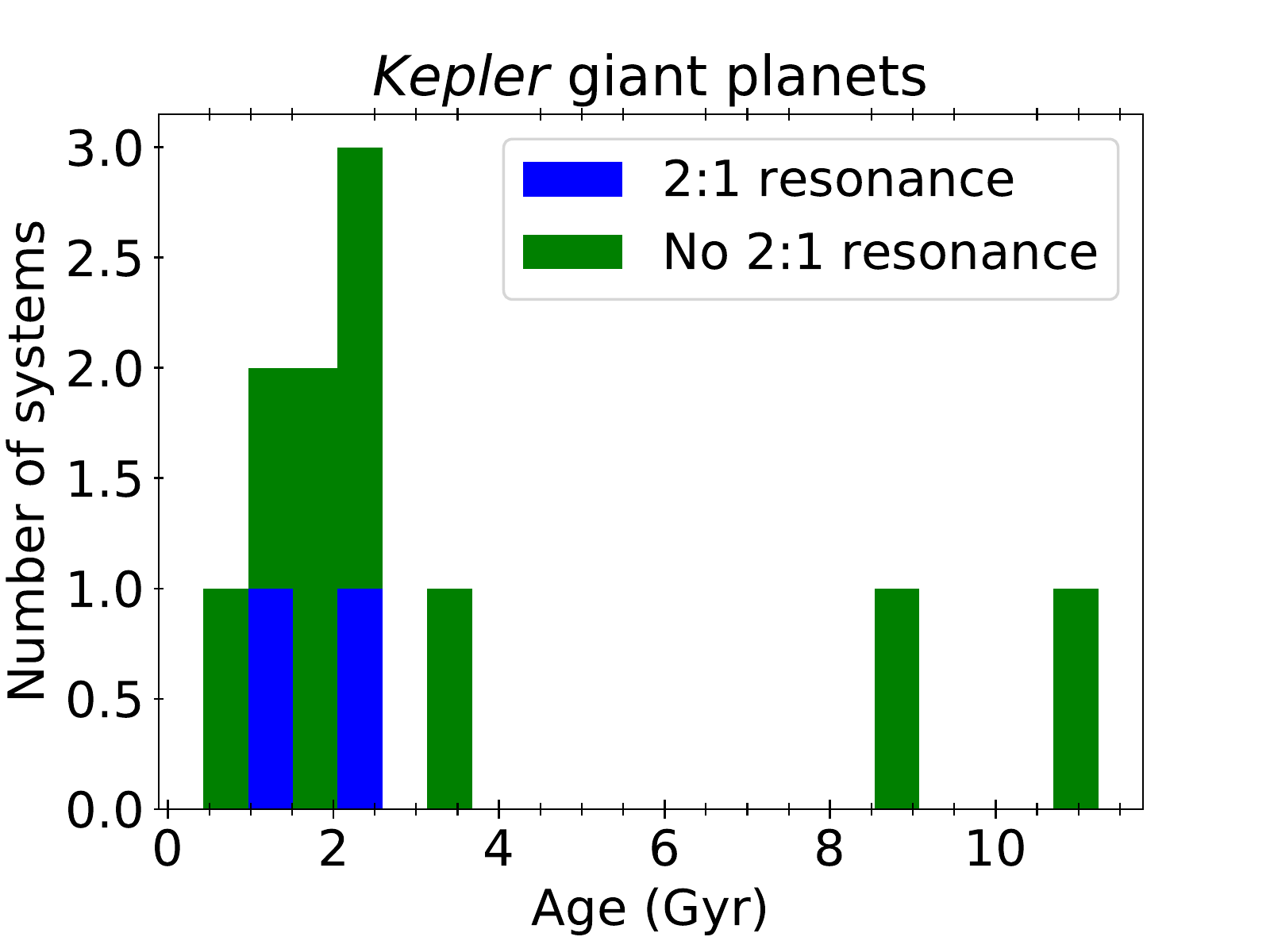}
    \includegraphics[width=3.5 in]{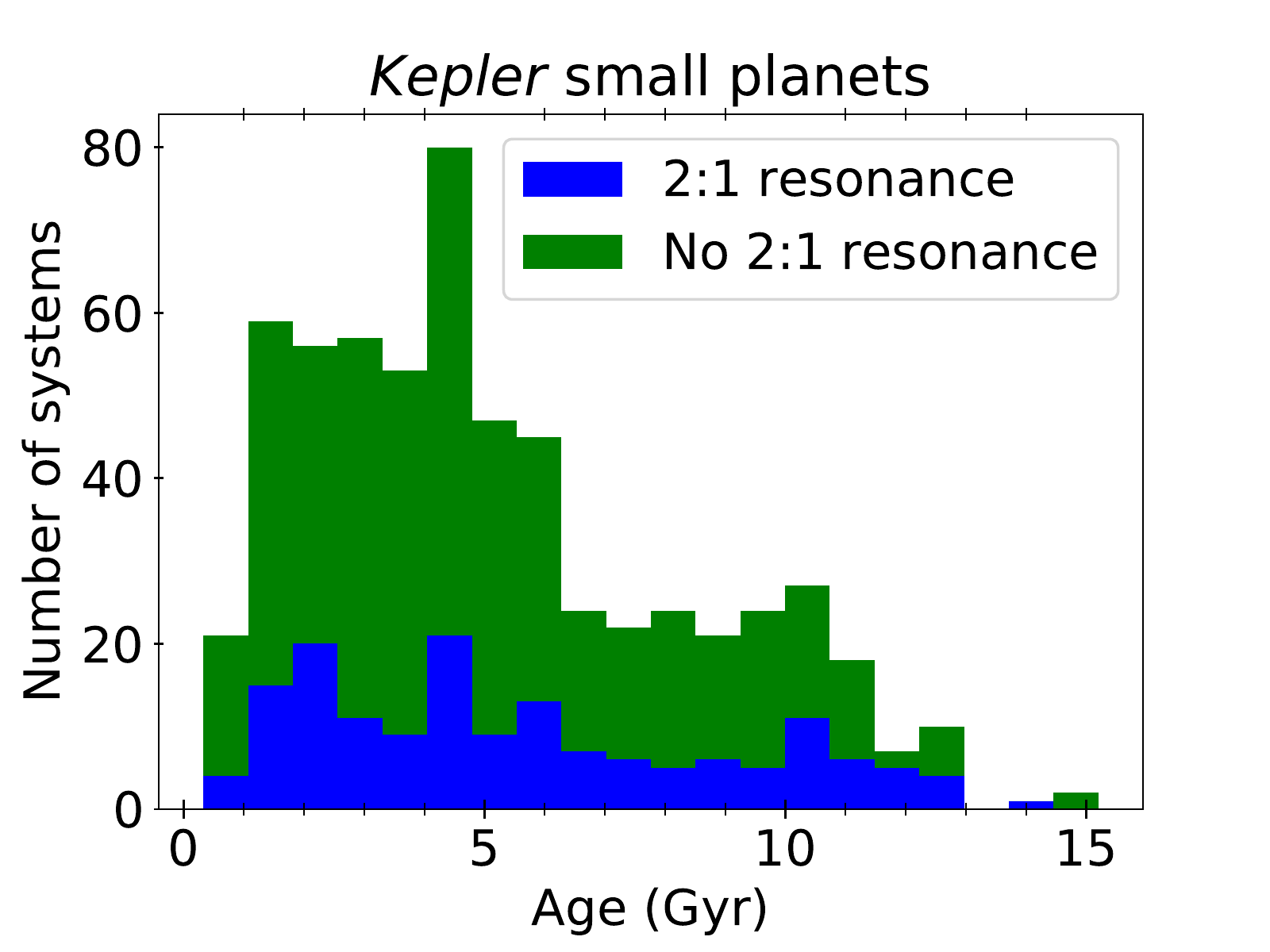}
    \caption{Ages histograms for the original \cite{kz2011} sample (upper left), massive planet pairs in the RV sample (upper right), giant planet pairs in the \textit{Kepler} sample (lower left), and small planet pairs in the \textit{Kepler} sample (lower right). Systems near a 2:1 resonance are shown in blue, and systems without a 2:1 resonance are shown in green.}
    \label{fig:resageshists}
\end{figure}

We apply Eqns. \ref{eqn:resnur} and \ref{eqn:resch} first to the original dataset of \cite{kz2011}, in order to compare with our original results.  A histogram of the ages in this sample is shown in the upper left panel of Figure \ref{fig:resageshists}.  There are 30 systems total; 5 are near a 2:1 resonance, and 25 are not.  The median age is 6.1 Gyr and the standard deviation is 1.8 Gyr.  The ages in this dataset were pulled from various surveys of Ca II H\&K lines and do not have individual uncertainties reported.  However, many of those surveys used a relation derived by \cite{mamhill2008}, who estimated an overall uncertainty of 60\% on ages derived from their relation when accounting for both calibration and observational uncertainties and astrophysical scatter.  Therefore, for each system, we assign $p(t_{\rm \star,obs}|t_\star)$ a Gaussian distribution with a width of 60\% of the reported age.  This analysis yields a ratio of the Nurture hypothesis to the Chance hypothesis of 2.2.  Within the significant figures we report, this value is the same as what we obtained without formally incorporating uncertainties, and is not strong enough for us to favor one hypothesis over the other.

We consider three additional updated samples.  First, we obtained data on systems discovered via radial velocity (RV) with pairs of adjacent massive planets (planetary mass $M_{\rm P}>0.3 M_{\rm J}$) and measured ages.  We acquired the data from NASA's Exoplanet Archive (exoplanetarchive.ipac.caltech.edu; \citealt{psrv}) on 2021 August 26, using values in the default parameter set, as well as data for systems whose only ages are not in the default parameter set. The total sample contains 40 planetary systems; of these, 7 have a planet pair near a 2:1 resonance, and 33 do not.  A histogram of stellar ages of 2:1 resonant and nonresonant systems in this sample is shown in the upper right panel of Figure \ref{fig:resageshists}.  The ages in this sample range between 0.5 and 10 Gyr, with a median age of 3.0 Gyr and standard deviation of 2.2 Gyr. Most ages have measurement uncertainties of $\sim$1-2 Gyr.  For those systems without reported age uncertainties, we take the median age uncertainty of the rest of the sample, which is 1.6 Gyr for the upper error and 1.5 Gyr for the lower error.  As this sample was pulled from the Exoplanet Archive, rather than a single study, the ages have been derived by a variety of methods, most commonly isochrones/evolutionary tracks or chromospheric activity.  This sample has a Nurture to Chance odds ratio of 2.0.

As a second updated sample, we use pairs of adjacent giant planets -- defined here to mean having planetary radius $R_{\rm P}>6R_\oplus$ -- discovered by \textit{Kepler}, with the ages calculated by \cite{berger2020} and planetary radii from \cite{berger2020planets}.  The planetary periods were obtained from the Exoplanet Archive on 2021 November 12 \citep{pskepler}. We use both candidate and confirmed systems (based on the disposition of the Exoplanet Archive) and exclude any that are not in the \cite{berger2020planets,berger2020} catalog.  \cite{berger2020} combined data from the \textit{Kepler} and \textit{Gaia} spacecraft and used isochrones to derive a homogeneous catalog of stellar ages and other properties.  They reported a median stellar age uncertainty of 56\%.  This sample contains 11 giant-planet pairs.  2 of these are near a 2:1 resonance, and 9 are not.  An ages histogram for this sample is shown in the lower left panel of Figure \ref{fig:resageshists}.  Most of the systems in this sparse sample are a couple Gyr old, with two older systems; the median age is 2.5 Gyr, with a standard deviation of 3.2 Gyr.  With this sample, we obtain an odds ratio of Nurture to Chance of 1.7.

Finally, we also consider pairs of adjacent small planets ($R_{\rm P}\leq 6 R_\oplus$) discovered by \textit{Kepler}, again with the \cite{berger2020} ages and \cite{berger2020planets} radii, and with the data on periods acquired from the Exoplanet Archive on 2021 November 12.  In this sample, there are 598 planetary systems.  158 of these are near a 2:1 resonance, and 440 are not.  An ages histogram for this sample is shown in the lower right panel of Figure \ref{fig:resageshists}.  The ages of both resonant and non-resonant systems in this sample range from about 0.5 to about 15 Gyr, with a median age of 4.5 Gyr and standard deviation of 3.1 Gyr. This sample yields an odds ratio of Nurture to Chance of 1.1.

\begin{figure}[ht]
    \includegraphics[width=3.5 in]{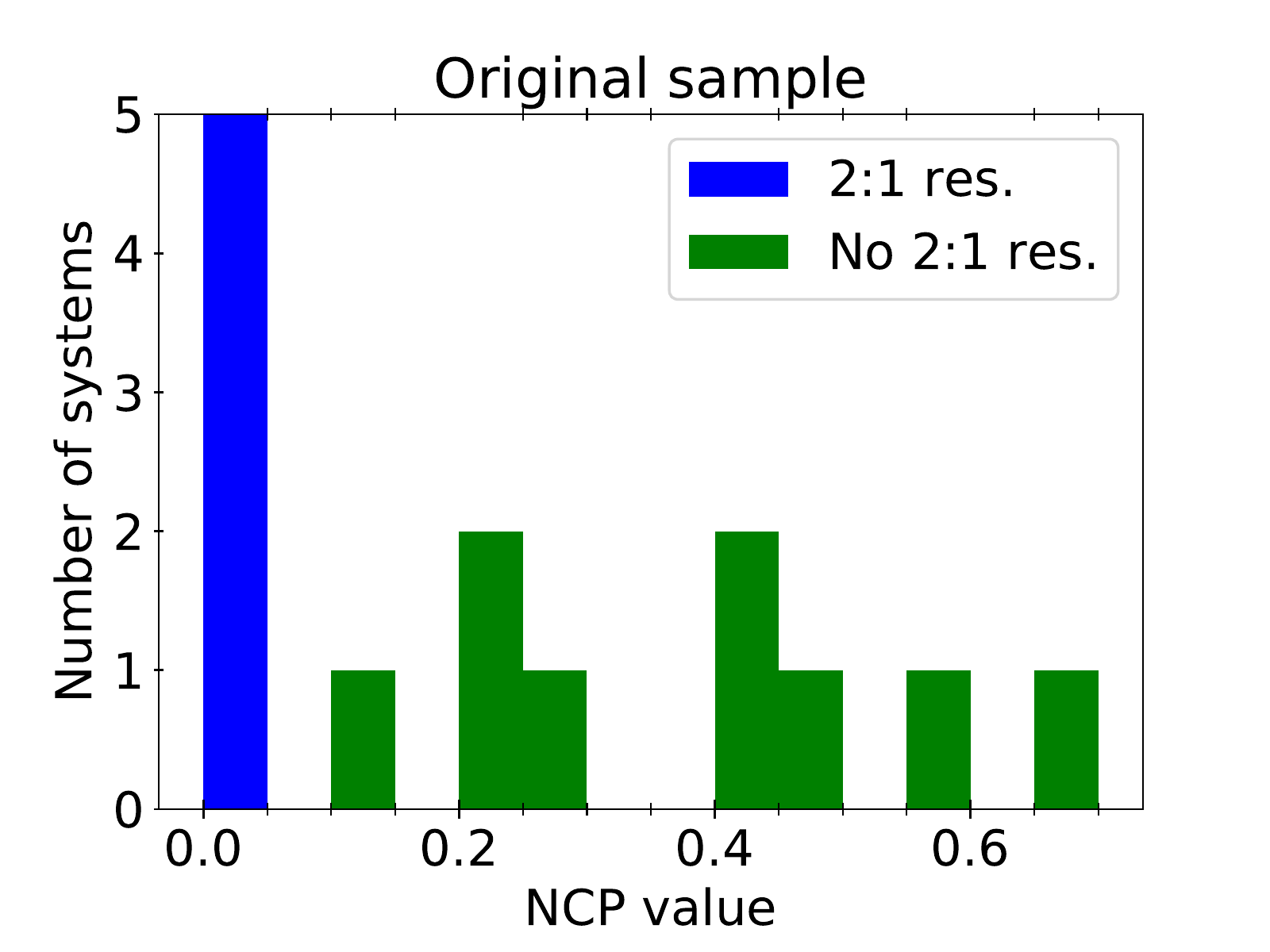}
    \includegraphics[width=3.5 in]{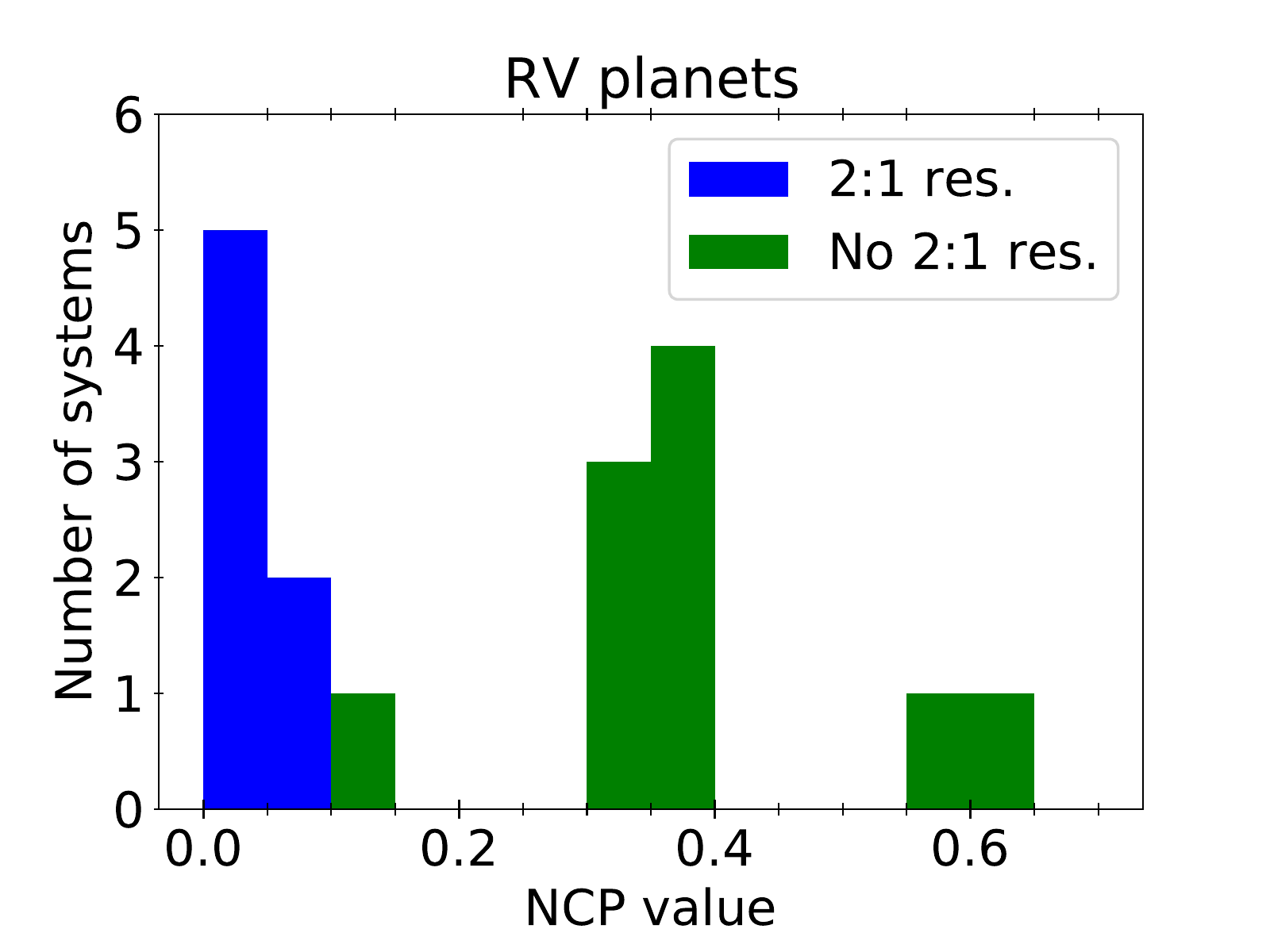}
    \includegraphics[width=3.5 in]{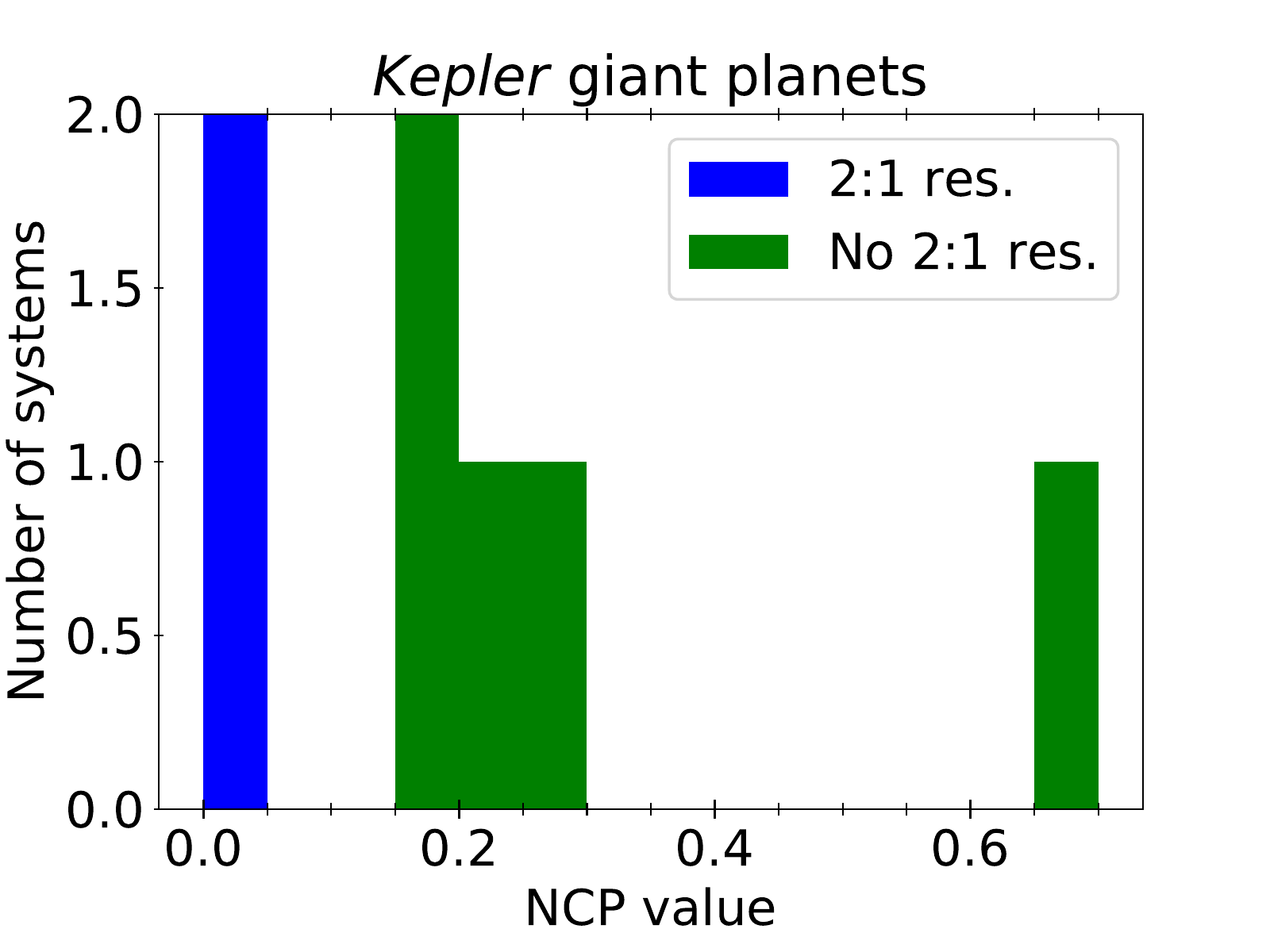}
    \includegraphics[width=3.5 in]{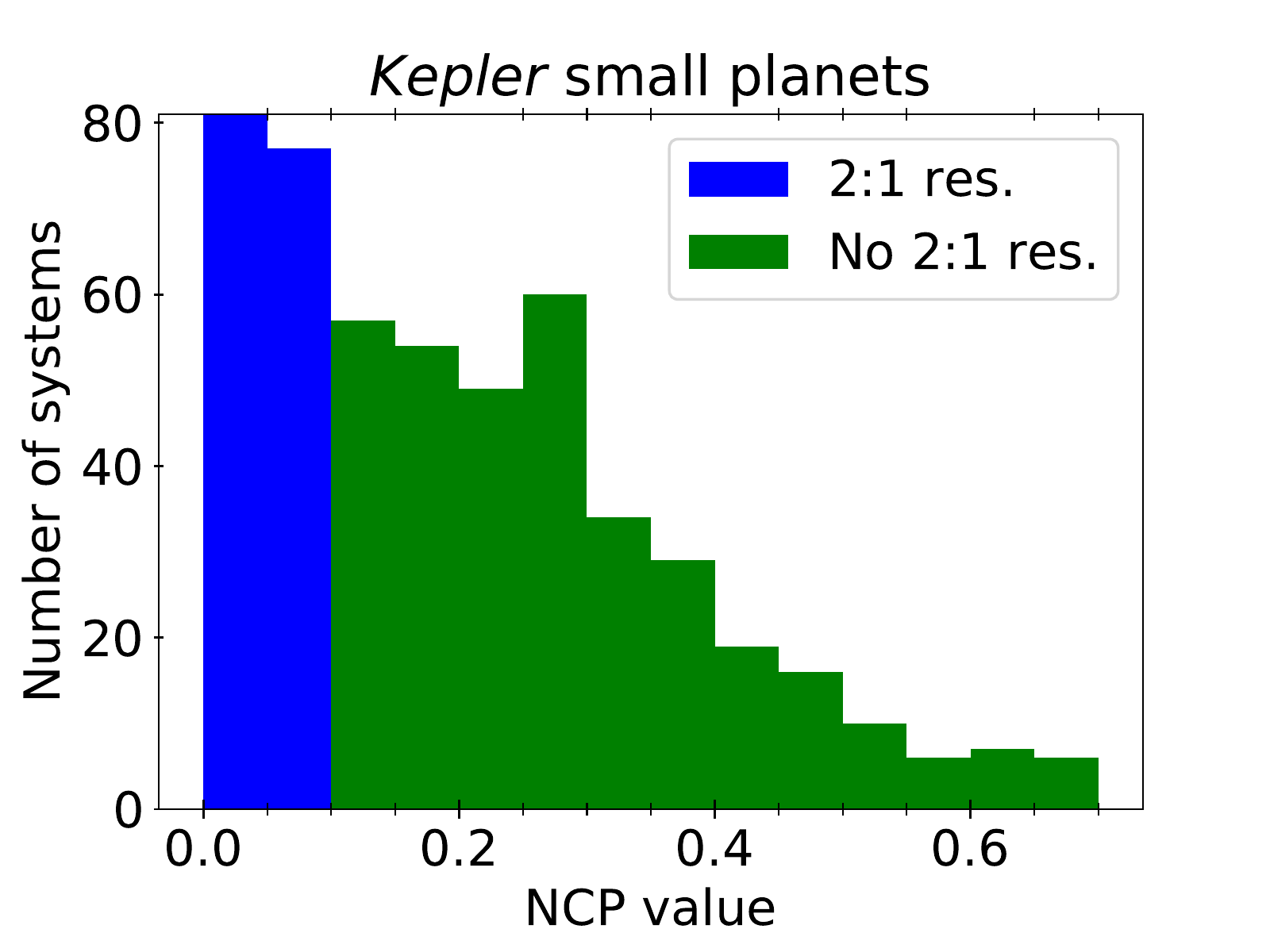}
    \caption{Histograms of NCP values for the original \cite{kz2011} sample (upper left), massive planet pairs in the RV sample (upper right), giant planet pairs in the \textit{Kepler} sample (lower left), and small planet pairs in the \textit{Kepler} sample (lower right). Systems near a 2:1 resonance are shown in blue, and systems without a 2:1 resonance are shown in green.  These plots are focused near $\delta=0.1$, the threshold for being near a 2:1 resonance, and do not show every system in the samples.}
    \label{fig:resratiohists}
\end{figure}

In Figure \ref{fig:resratiohists}, we display histograms of the NCP values for the original \cite{kz2011} sample and for each of the three updated samples.  These plots do not show every system in the samples, but instead are focused near $\delta=0.1$, the threshold for being near a 2:1 resonance.  In the original sample, there is one system close to the resonance threshold, but otherwise the 2:1 resonant and nonresonant systems are well separated.  There is one nonresonant RV system near the threshold, but otherwise, for the RV and \textit{Kepler} giant planet samples, there is a noticeable separation between 2:1 resonant and nonresonant systems.  There is less of a separation for the \textit{Kepler} small planets.

Table \ref{tab:resoddsratios} compares the odds ratios obtained without (from Paper 1) and with incorporating uncertainties, for the original sample as well as the updated samples.  None of the odds ratios for our updated samples are different enough from our original result to allow us to favor one hypothesis over another.  In other words, the relation between 2:1 resonances and age that we have modeled here is just as good as the ages of both 2:1 resonant and nonresonant systems being determined purely by chance.  One reason for this comparability is that the ranges we use for our hyperpriors are designed to include a wide range of possibilities for the 2:1 resonance disruption timescale.  This means that we potentially include both high-probability and low-probability regions in the $\mu-\sigma$ parameter space.  Further testing shows that restricting, for example, $\mu$ to (6,7) and $\sigma$ to (0,1) could produce odds ratios that strongly favor the Chance hypothesis (these ranges were chosen to span relatively low-probability regions of the plots shown in Figure 5 in Paper 1, which show probability contour plots for the hyperparameters $f_0$, $\mu$, and $\sigma$). In such a scenario, the majority of 2:1 resonances get disrupted very early on, so the presence of a 2:1 resonance in a system a few Gyr old goes against the proposed age trend, and thus Chance is favored.  However, our lack of knowledge of resonance disruption timescales and the possibility of a wide range of timescales (e.g. \citealt{izid17}) prevents us from making any meaningful restrictions on hyperprior ranges.  If there is a large variation in the disruption timescale from system to system (i.e. a high value of $\sigma$), an age trend may be inherently difficult to distinguish from Chance, even with a large sample such as that of the \textit{Kepler} small planets.  We discuss this more in the next section.

\begin{table}
	\centering
	\begin{tabular}{l|l|llll}
	    \multicolumn{6}{c}{2:1 ORBITAL RESONANCES} \\
		\hline
		& Without uncertainties& 
			    \multicolumn{4}{c}{With uncertainties} \\
		Ratio & \cite{kz2011} & \cite{kz2011} & RV & Kepler Giant & Kepler Small \\
		\hline\hline
        $p(H_{\rm nur})/p(H_{\rm ch})$ & 2.2 & 2.2 & 2.0 & 1.7 & 1.1\\
	    \hline
	\end{tabular}
	\caption{Odds ratios for the 2:1 resonances cases without (from Paper 1) and with incorporating measurement uncertainties, for the original sample as well as the updated samples.}
	\label{tab:resoddsratios}
\end{table}

\subsection{Simulated Resonance Data}
In this section, we investigate the types of age trends we might be able to detect with available and future age datasets, as well as the effects of measurement uncertainties on the odds ratios.  We begin by using the original ages used by \cite{kz2011} and assume these are the true ages of the stars.  There are 30 stars in this sample.  We choose a value for $\mu$ and a value for $\sigma$, to define the underlying population of $t_{\rm d}$ values.  For each stellar age, we draw a value for $t_{\rm d}$ from the log-normal distribution set by $\mu$ and $\sigma$.  If the selected value of $t_{\rm d}$ is greater than the age, the system is classified as being near a 2:1 resonance; otherwise, it is classified as non-resonant.  This first experiment assumes that $f_0=1$, i.e. that all systems begin with a 2:1 resonance.  We then simulate measurements of each stellar age by randomly drawing from a normal distribution centered at the true age and with a width of an assumed measurement precision. We do this 100 times for the whole dataset, and calculate two versions of the Nurture to Chance odds ratio for each set of observations, one accounting for measurement uncertainties and one without accounting for measurement uncertainties.  We perform these calculations for a range of values of $\mu$ and $\sigma$ at two different levels of measurement uncertainty, 20\% and 60\% of the stellar age.  The latter value is comparable to the median age uncertainties in the \cite{kz2011} and \cite{berger2020} datasets.

We test values of $\mu=9$, 9.5, and 9.75 -- mean disruption timescales comparable to the ages in the sample -- and $\sigma=0.1$, 1, and 5.  In all these cases, the odds ratios are of order unity, regardless of the level of measurement uncertainty and regardless of whether the uncertainties are marginalized over or not.  Using an initial fraction of $f_0=0.5$ or 0.35, instead of 1, makes no significant difference.  With the uninformative prior and hyperpriors we have chosen to describe the 2:1 resonance disruption timescale, values of $\mu$ comparable to system ages are the most likely to yield strong odds ratios in favor of Nurture; since, with this dataset, they do not, we do not test other values of $\mu$.

Next, we perform this exercise using a somewhat larger artificial sample of 45 stellar ages randomly selected between 0 and 13.7 Gyr, and test $\mu$ values of 7, 9, 9.5, 10, and 11, and $\sigma$ values of 0.01, 0.1, 1, and 5.  For most values of $\mu$ and $\sigma$, we obtain odds ratios of order unity, indicating nearly equal support for the Nurture and Chance hypotheses in the simulated data.  For simulated sets with a value of $\mu$ significantly higher or lower than the age range in the data, the comparable support for each hypothesis may be due to the limited age range and size of the simulated data.  If the typical disruption timescale is, for instance, on the order of 100 Gyr, a dataset with a maximum age of 13.7 Gyr will have most systems still in their initial resonance state.  This scenario is essentially equivalent to Chance, because in both situations, the systems are observed to have the same resonance state with which they formed.  Similarly, if the typical disruption timescale is on the order of 10 Myr, a dataset uniformly spread between 0 and 13.7 Gyr will have most initially resonant systems disrupted, and would need many age measurements of very young stars for an age trend to be evident.  Thus, an age trend in these scenarios is inherently difficult to distinguish from a Chance relation, given the limitations of a typical age dataset, because they do not show a transition in the number of 2:1 resonant systems with age.

\begin{figure}[ht]
    \includegraphics[width=3.5 in]{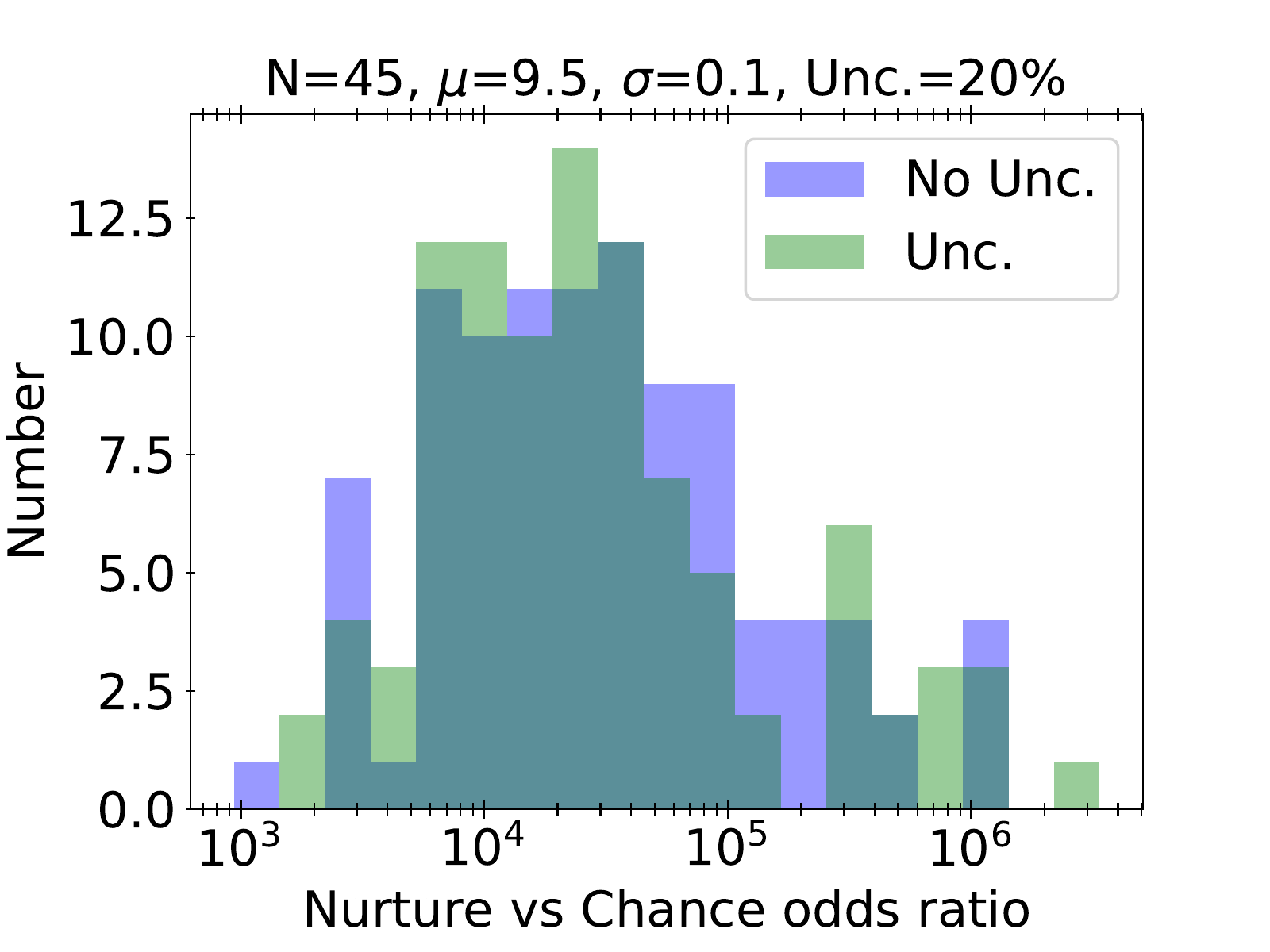}
    \includegraphics[width=3.5 in]{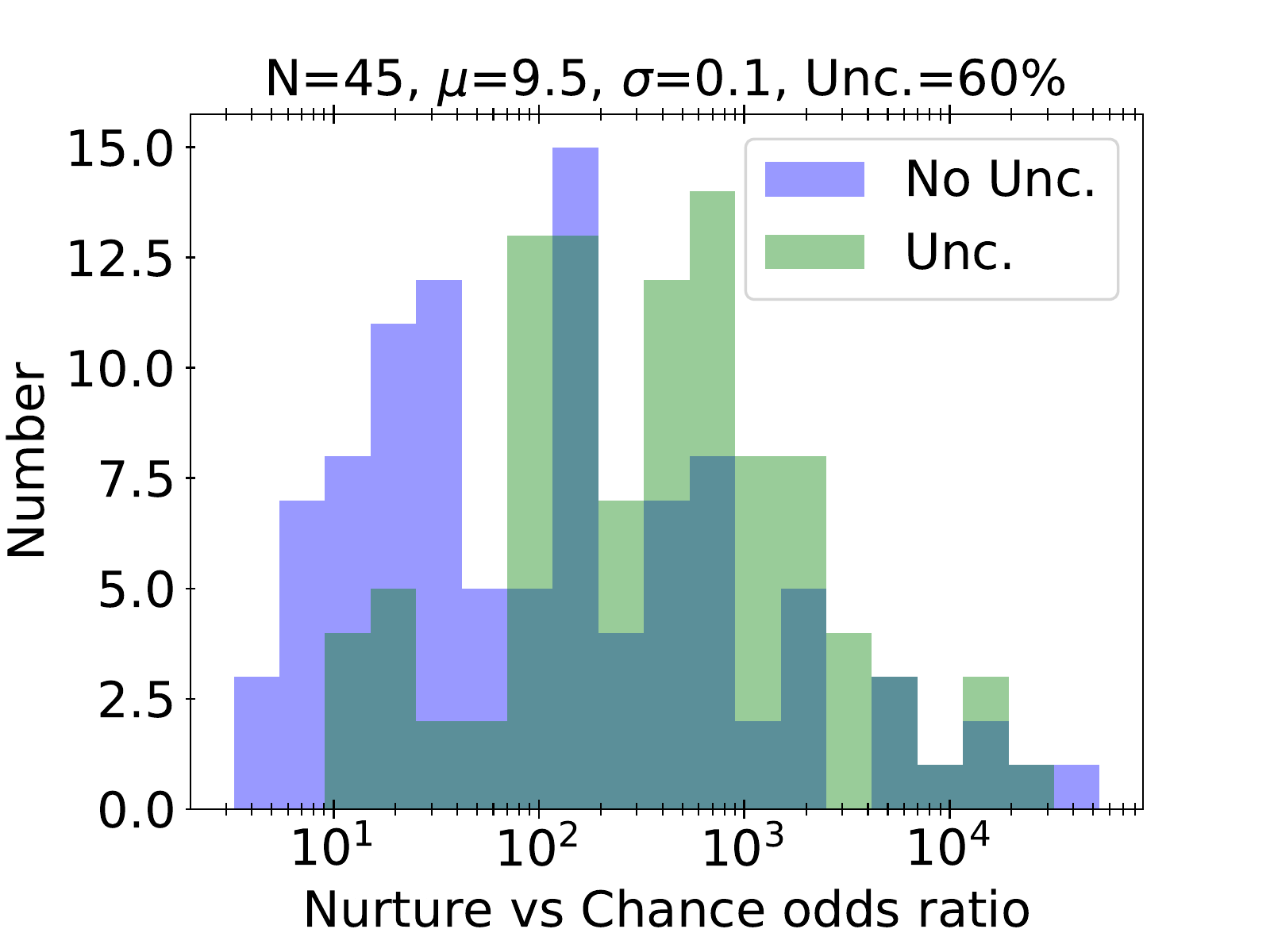}
    \includegraphics[width=3.5 in]{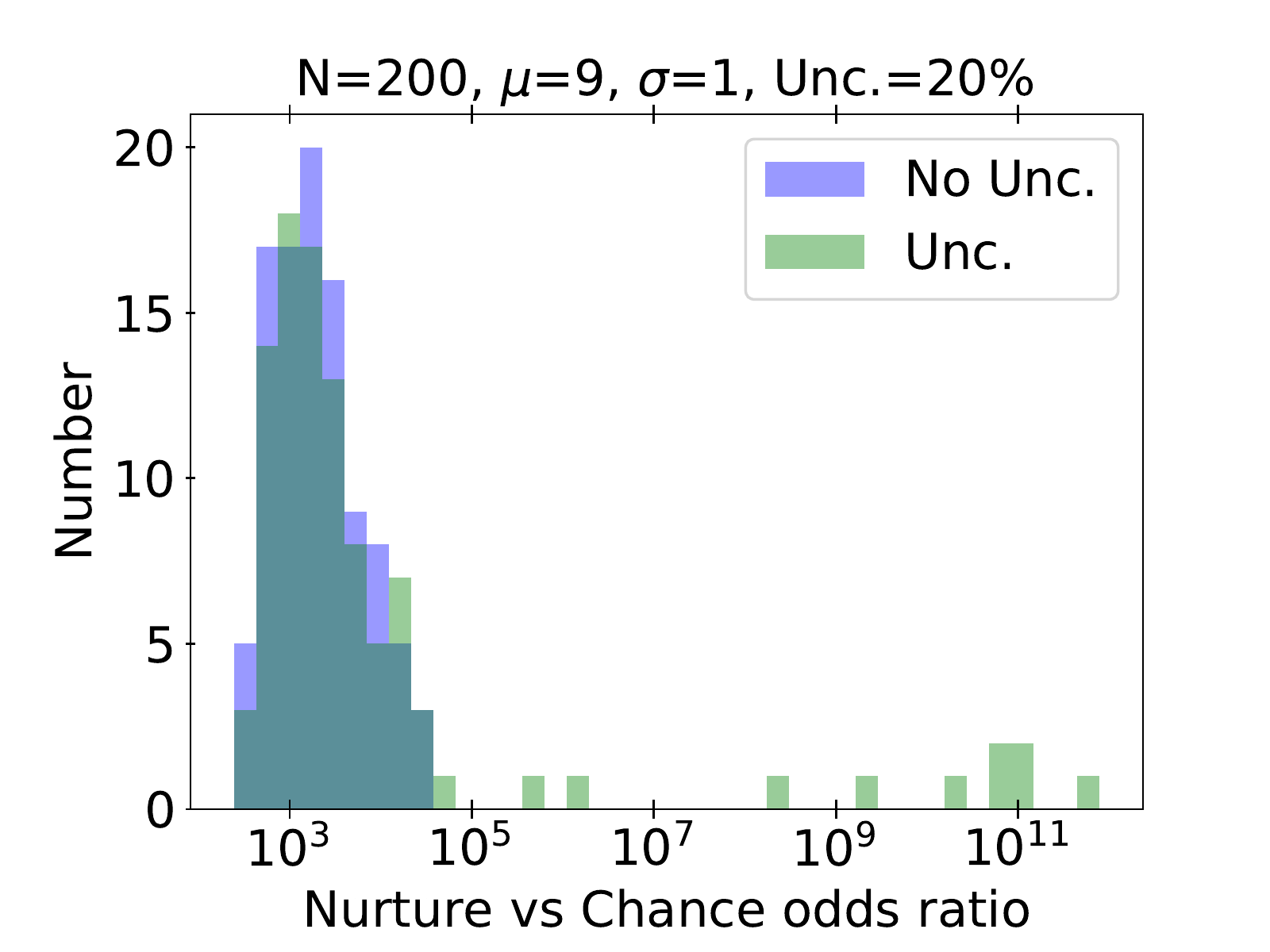}
    \includegraphics[width=3.5 in]{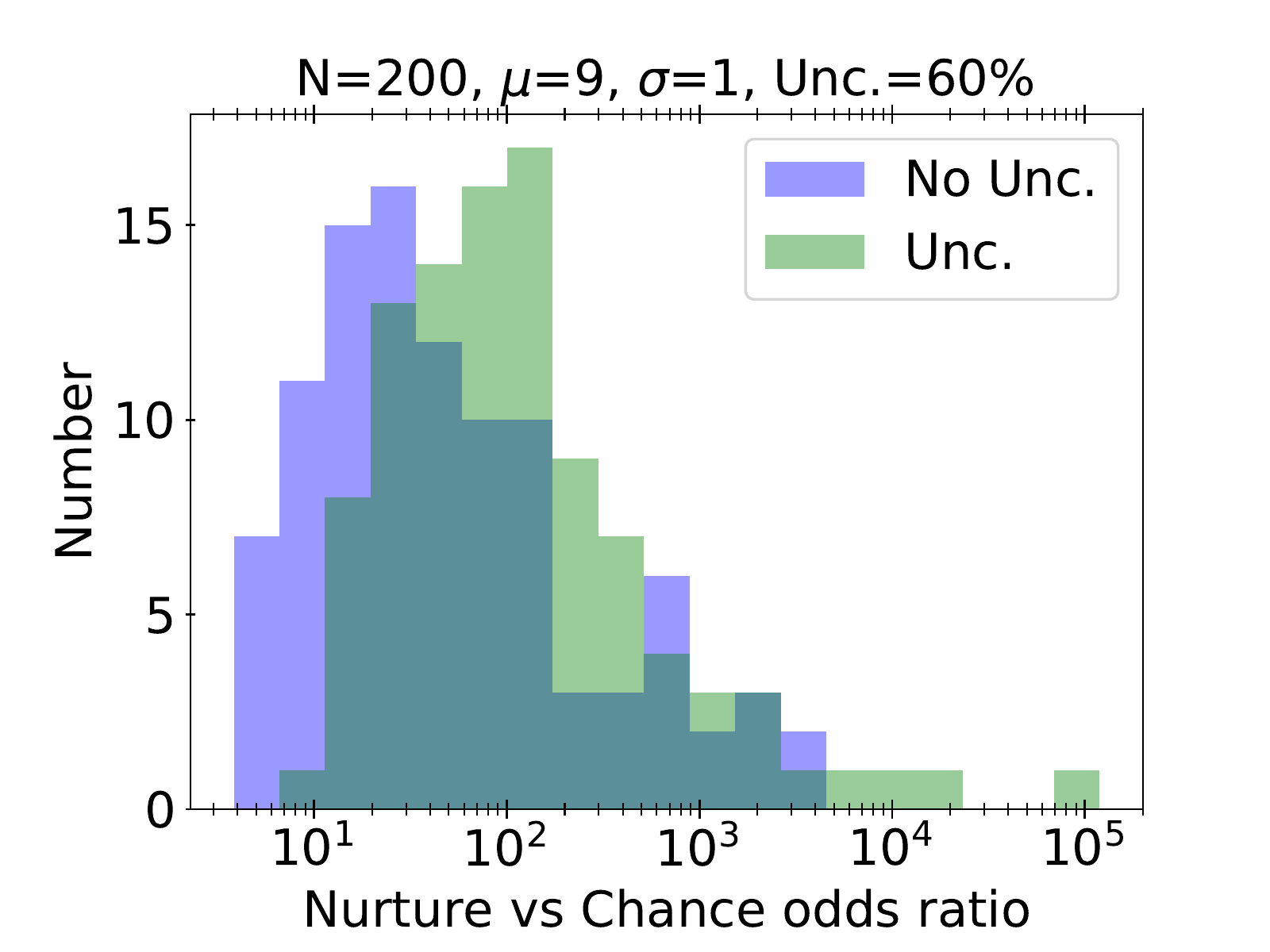}
    \includegraphics[width=3.5 in]{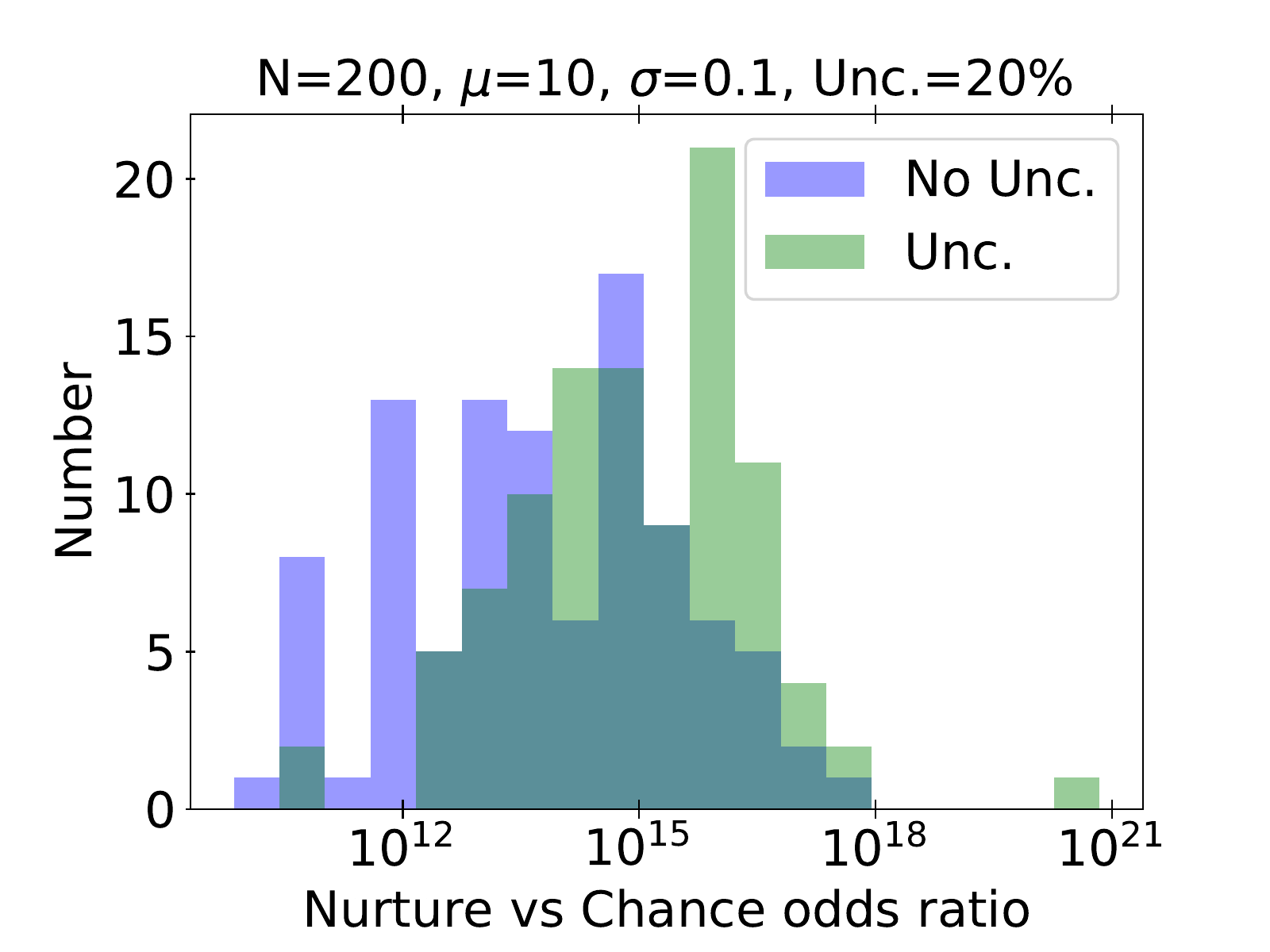}
    \includegraphics[width=3.5 in]{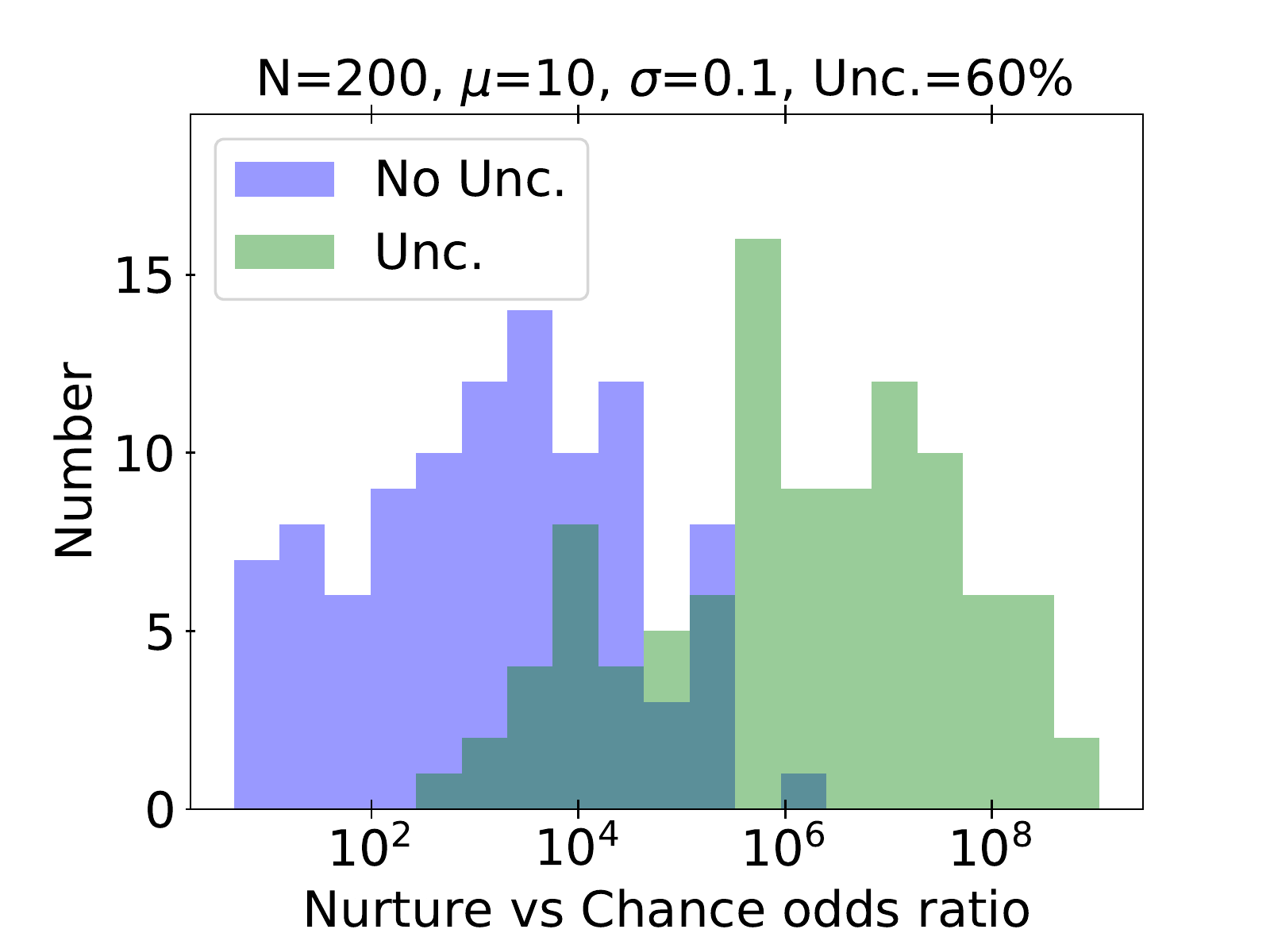}
    \caption{Distributions of Nurture vs. Chance odds ratios for various values of simulated dataset size $N$, $\mu$, $\sigma$, and measurement uncertainty, generated from 100 ``observations" of the true simulated ages.  Ratios calculated without incorporating measurement uncertainties are shown in blue, and ratios calculated with incorporating uncertainties are shown in green. Note that even when uncertainties are not accounted for in the odds ratio, the level of uncertainty does affect the observed ages because we don't perfectly measure the true quantities.  This is why the blue histograms differ between the left and right panels.}
    \label{fig:resfakedata}
\end{figure}

Of the values for $\mu$ and $\sigma$ that we tried, the support for the Nurture hypothesis over Chance was greatest with $\mu=9.5$ and $\sigma=0.01$ or 0.1, i.e. a scenario in which 2:1 resonances typically get disrupted after a few billion years, and there is little variation in the disruption timescale from system to system.  Histograms of the distribution of Nurture vs. Chance odds ratios for $\mu=9.5$ and $\sigma=0.1$ are shown in the top row of Figure \ref{fig:resfakedata}.  These plots show the ratios for both levels of measurement uncertainty, and for the calculations performed without incorporating uncertainties (in blue) and with incorporating uncertainties (in green) into the odds ratios. Note that even when uncertainties are not accounted for in the odds ratio, the level of uncertainty does affect the observed ages because we don't perfectly measure the true quantities.  This is why the blue histograms differ between the left and right panels.  With 20\% measurement uncertainties, the odds ratios span roughly three orders of magnitude, but all strongly favor the Nurture hypothesis.  Further, there is strong overlap between the blue and green histograms, suggesting that in this case, incorporating uncertainties has little effect on the odds ratios.  With 60\% measurement uncertainties, the ratios again span several orders of magnitude, but are, on the whole, lower than the ratios from 20\% uncertainties.  Without incorporating uncertainties, roughly half of the odds ratios are below the threshold of 100 for strongly favoring Nurture.  When uncertainties are incorporated, the distribution shifts somewhat to higher ratios, though there is still a good amount of overlap between the blue and green histograms.  This suggests that incorporating uncertainties may have a somewhat stronger effect on the 60\% age uncertainty odds ratios compared to the 20\% age uncertainty odds ratios.  The Nurture hypothesis is also strongly favored with $\mu=10$ and $\sigma=0.01$ when the ages are measured with 20\% uncertainty, and again we see significant overlap between the distributions with and without incorporating uncertainties (though this scenario is not shown in our plots), but with 60\% uncertainties, the odds ratios are all inconclusive.  For both $\mu=9.5$ and $\mu=10$, increasing $\sigma$ by an order of magnitude yields only odds ratios that do not strongly favor either hypothesis.

We perform this same exercise again using a large simulated dataset, consisting of 200 stellar ages randomly drawn between 0 and 13.7 Gyr.  We test $\mu=7$, 9, 9.5, 10, and 11, and $\sigma=0.1$, 1, and 5.  With a larger sample, a disruption timescale distribution with $\sigma=5$ still does not yield conclusive odds ratios, and, for values of $\mu$ significantly smaller or larger than the typical stellar age, neither does $\sigma=1$.  However, this larger sample strengthens the odds ratios for several other values of $\mu$ and $\sigma$, particularly for $\mu=9$ or 9.5.  We display a few of the cases where Nurture is strongly favored in the middle and bottom rows of Figure \ref{fig:resfakedata}.  The middle row shows $\mu=9$ and $\sigma=1$.  The Nurture hypothesis is strongly favored for all observations for 20\% measurement uncertainty, regardless of whether the uncertainties are marginalized over or not.  When uncertainties are incorporated, there are a few very high, outlying odds ratios, but the bulk of the distribution strongly overlaps with the no-uncertainties version.  Zooming in on the bulk of the distribution makes this overlap more apparent and shows no significant shift between the uncertainties and no-uncertainties distributions.  The bottom left panel shows that Nurture is also overwhelmingly favored with $\mu=10$, $\sigma=0.1$, and 20\% age uncertainty.  The green distribution again overlaps strongly with the blue, though it is shifted to higher values.

In general, the odds ratios calculated with 60\% age uncertainties in the larger simulated dataset tend to be lower than those with 20\% uncertainties, but usually result in the same overall conclusion. An exception to this is when $\mu=9$ and $\sigma=1$, shown in the middle row of Figure \ref{fig:resfakedata}.  With 20\% uncertainties, Nurture is very strongly favored, but with 60\%, the median odds ratio falls below the threshold of 100, whether uncertainties are incorporated or not.  With 60\% uncertainties, incorporating uncertainties usually doesn't affect the overall conclusion.  There is usually significant overlap between the uncertainties and no-uncertainties odds ratio distributions.  However, for 60\% uncertainties, there is often a noticeable upward shift in the distribution when uncertainties are incorporated.  Within the parameters we tested, this shift is most pronounced for $\mu=10$ and $\sigma=0.1$, shown in the bottom right of Figure \ref{fig:resfakedata}.

With the large simulated dataset, at least within the parameters we tested, the level of measurement uncertainty, and whether uncertainties are marginalized over or not, does not usually have an effect on the overall conclusions from the odds ratio calculations.  For ratios that already strongly favor Nurture, marginalizing over uncertainties may affect the value of the ratio by even several orders of magnitude, but Nurture is usually still strongly favored.  On the other hand, the closer the median ratio is to 1, the smaller the effect of measurement uncertainties, and the smaller the spread of ratios from the different simulated observations.

These results indicate that most of the datasets currently available are inadequate for assessing most possible age trends.  It is possible that a trend of 2:1 resonances with age may yet exist, but is obscured by measurement uncertainty.  More precise stellar age measurements may help settle this question.  However, this exercise also suggests that even if 2:1 resonances get disrupted over time, the limited range of a typical stellar age dataset, even with well-measured ages, may preclude us from favoring an age trend over Chance.  If the typical 2:1 resonance disruption timescale is much younger or much older than a few Gyr, or if disruption happens on a wide variety of timescales (i.e. if $\sigma\gtrsim1$) as some studies (e.g. \citealt{izid17}) indicate, these situations may be difficult to distinguish from a Chance relation because the available ages do not span enough orders of magnitude.  Age measurements of very young multiplanet systems may help this issue if the typical 2:1 resonance disruption timescale is relatively short.

\section{Obliquities of Stars with Hot Jupiters}
\label{section:obliquities}
Stellar obliquity refers to the tilt of a star's rotation axis relative to the orbital axis of its planet.  Obliquities may be excited or dampened by a number of mechanisms, such as planet-planet scattering or tidal interactions; see Section 3.2 of \cite{hotjupreview2018} for a review.  Understanding the dominant processes that control stellar obliquities will shed light on the formation and orbital evolution of their associated planets.

\cite{winn2010} found a correlation between the effective temperatures and obliquities of stars with hot Jupiters: hot stars have high obliquities, and cool stars have low obliquities.  They separated ``hot" and ``cool" stars at a stellar effective temperature of $T_{\rm eff,\star}=6250$ K, near the Kraft break \citep{kraft1967}.  They proposed that all stars start out with a wide range of obliquities, and the cooler stars quickly realign with planetary orbits due to their large convective envelopes; hotter stars, without such convective envelopes, remain misaligned.  This trend with temperature was also found by \cite{schlaufman2010} using projected rotational velocity measurements and confirmed by  \cite{albrecht2012} with a much larger sample of obliquities.  On the other hand, \cite{triaud2011} described a trend of stellar obliquity with age, where young stars have a wide range of obliquities but older stars are aligned.  They noted that stars more massive than $1.2 M_{\odot}$ will cool considerably over their main sequence lifetimes.  This evolution means, they proposed, that the cool stars with low obliquities could be older stars that slowly tidally realigned as they aged and cooled down.  This connection between age and temperature can make it difficult to tell which is the fundamental driving force behind the distribution of stellar obliquities.

In Paper 1, we used our Bayesian framework to determine whether the sample from \cite{triaud2011} better supported a model of gradual tidal realignment over time (Nurture hypothesis), or a model of hot stars having high obliquities and cool stars being well aligned (Nature hypothesis).  This dataset consisted of 22 stars with mass $M_{\star}>1.2 M_{\odot}$.  A star was deemed misaligned if it had a measured projected obliquity angle $\lambda > 20^\circ$; under this criterion, the sample contained 10 misaligned stars and 12 aligned stars \citep{triaud2011}.  We note that while cool stars do get realigned in the temperature-dependent scenario, the timescale for the realignment would need to be short enough, and occur soon enough after the planet's formation or arrival close to the star, to produce such a sharp trend that we are comfortable classifying this scenario as the Nature hypothesis.  We found strong support for the Nature hypothesis, with an odds ratio for Nature vs. Nurture of 210.  This conclusion held up under additional tests, including removing stars with highly uncertain ages, and bootstrapping the data. Here, we reanalyze this dataset as well as an updated sample with our modified framework to formally incorporate measurement uncertainties.

For simplicity, and to keep in line with \cite{winn2010} and \cite{triaud2011}, rather than using the actual value of a system's measured obliquity, we consider whether the obliquity means the system is aligned or misaligned.  We represent this property as the binary parameter $A$, where $A=1$ if a system is misaligned and $A=0$ if a system is aligned.  $A_{\rm obs}$ represents the observed alignment state of a system, and $A$ represents the true alignment state of a system.  The term $p(A_{\rm obs}|A)$ accounts for the uncertainty in the measured projected obliquity, specifically uncertainty as to whether the system is aligned or misaligned.  To account for this, we impose the criterion that a system is considered misaligned if it has $\lambda>10^\circ$ at the $3\sigma$ level.  This is according to \cite{winn2010} and is a departure from the $\lambda>20^\circ$ criterion used by \cite{triaud2011} and which we used in Paper 1.  However, for the \cite{triaud2011} sample, these two criteria yield the same sets of aligned and misaligned systems.  For our updated sample, the new criterion means that the misaligned group will only contain stars with obliquities measured well enough to be confidently considered misaligned.  Accordingly, we assume $A_{\rm obs}=A$ and drop the $p(A_{\rm obs}|A)$ term from the following equations.

Since a primary concern in this particular application is the evolution of stellar temperature with age -- which makes the true source of the obliquity trend more difficult to distinguish -- we need a way to account for this interdependence in the joint distribution $p(t_\star,T_{\rm eff,\star})$. 
To do this, we use the {\tt isochrones} Python package \citep{isochrones}.  As input to this program, we use each sample star's $T_{\rm eff,\star,obs}$, log($g$)$_{\rm obs}$, [Fe/H]$_{\rm obs}$, and the associated uncertainties for these parameters as listed in the TEPCat catalog \citep{tepcat}, as of 2021 April 29.  The program output is a joint distribution of stellar age and stellar effective temperature, which we normalize to have an area of 1 and which we will refer to in the following equations as $p(t_{\rm \star,iso},T_{\rm eff,\star,iso})$.  We will then marginalize over $t_{\rm \star,iso}$ and $T_{\rm eff,\star,iso}$.  The likelihood equations adapted for each hypothesis in this application are as follows.

In the Nurture (time-evolution) hypothesis, we assume that $A$ can only evolve from 1 to 0, i.e. systems can go from misaligned to aligned, but not the other way around.  We introduce the alignment timescale $t_{\rm a}$, the time it takes a system to go from $A=1$ to $A=0$.  We assume the underlying population of $t_{\rm a}$ has a lognormal distribution with mean $\mu$ and standard deviation $\sigma$.  Both $\mu$ and $\sigma$ are hyperparameters that we give uniform hyperpriors, in log$_{10}$[yr] space, of $p(\mu)=U(6,16)$ and $p(\sigma)=U(0,20)$, respectively.  We also include the hyperparameter $f_0$, the fraction of systems that form misaligned, and give this a uniform hyperprior from 0 to 1.  Then the equation for the Nurture hypothesis is (derived from Eqn. \ref{eqn:gennurturebinary} and analogous to Eqn. 27 in Paper 1):
\begin{align}
   p(A_{\rm obs},&T_{\rm eff,\star,obs},\text{log}(g)_{\rm obs},\text{[Fe/H]}_{\rm obs}|\mu,\sigma,f_0)\nonumber\\&=
   \begin{cases}
   \iint p(T_{\rm eff,\star,obs},\text{log}(g)_{\rm obs},\text{[Fe/H]}_{\rm obs}) p(T_{\rm eff,\star,iso},t_{\rm \star, iso}|T_{\rm eff,\star,obs},\text{log}(g)_{\rm obs},\text{[Fe/H]}_{\rm obs}) \\ \times \left[(1-f_{0})+f_{0}\int_0^{t_{\rm \star,iso}} p(t_{\rm a}|\mu,\sigma) dt_{\rm a}\right] dT_{\rm eff,\star,iso} dt_{\rm \star, iso} &, A_{\rm obs}=0 \\
   \iint p(T_{\rm eff,\star,obs},\text{log}(g)_{\rm obs},\text{[Fe/H]}_{\rm obs}) p(T_{\rm eff,\star,iso},t_{\rm \star, iso}|T_{\rm eff,\star,obs},\text{log}(g)_{\rm obs},\text{[Fe/H]}_{\rm obs}) \\ \times \left[f_{0} \int_{t_{\rm \star,iso}}^\infty p(t_{\rm a}|\mu,\sigma) dt_{\rm a}\right] dT_{\rm eff,\star,iso} dt_{\rm \star, iso} &, A_{\rm obs}=1.
   \end{cases}
\end{align}

The Nature hypothesis for the alignment case says that stars above $T_{\rm eff,\star}=6250$K may be misaligned, while stars cooler than this must be aligned.  This leads us to use the following for $p(A|T_{\rm eff,\star})$:
\begin{align}
    p(A|T_{\rm eff,\star}<6250\text{ K},f_{\rm h})&=
    \begin{cases}
    1 &,A_{\rm obs}=0\\
    0 &,A_{\rm obs}=1
    \end{cases}\\
    p(A|T_{\rm eff,\star}\geq6250\text{ K},f_{\rm h})&=
    \begin{cases}
    1-f_{\rm h}&,A_{\rm obs}=0\\
    f_{\rm h}&,A_{\rm obs}=1.
    \end{cases}
\end{align}
Here, $f_{\rm h}$ is a hyperparameter representing the overall fraction of hot stars that are misaligned.  We assign $f_{\rm h}$ a uniform hyperprior from 0 to 1.  Then the equation for the Nature hypothesis is (derived from Eqn. \ref{eqn:gennature} and analogous to Eqn. 28 in Paper 1):
\begin{align}
\label{eqn:alignnat}
p(A_{\rm obs},&T_{\rm eff,\star,obs},\text{log}(g)_{\rm obs},\text{[Fe/H]}_{\rm obs}|f_{\rm h}) =\iint p(T_{\rm eff,\star,obs},\text{log}(g)_{\rm obs},\text{[Fe/H]}_{\rm obs}) \nonumber\\ &\times p(T_{\rm eff,\star,iso},t_{\rm \star, iso}|T_{\rm eff,\star,obs},\text{log}(g)_{\rm obs},\text{[Fe/H]}_{\rm obs}) p(A|T_{\rm eff,\star,iso},f_{\rm h}) dT_{\rm eff,\star,iso} dt_{\rm \star,iso}.
\end{align}

In the Chance hypothesis, there is no dependence of stellar obliquity on stellar temperature or age.  The only thing that determines whether a system is aligned or misaligned is the hyperparameter $f$, the overall fraction of systems that are misaligned.  We give $f$ a uniform hyperpior from 0 to 1.  Then the equation for the Chance hypothesis is (derived from Eqn. \ref{eqn:genchancebinary} and analogous to Eqn. 31 in Paper 1):
\begin{align}
    \label{eqn:alignchancev2}
    p(A_{\rm obs},&\text{log}(g)_{\rm obs},T_{\rm eff,\star,obs},\text{[Fe/H]}_{\rm obs}|f)\nonumber\\&=
    \begin{cases}
    (1-f) \iint p(T_{\rm eff,\star,obs},\text{log}(g)_{\rm obs},\text{[Fe/H]}_{\rm obs})  \\ \times  p(T_{\rm eff,\star,iso},t_{\rm \star, iso}|T_{\rm eff,\star,obs},\text{log}(g)_{\rm obs},\text{[Fe/H]}_{\rm obs}) dT_{\rm eff,\star,iso} dt_{\rm \star,iso} &, A_{\rm obs}=0\\
    f \iint p(T_{\rm eff,\star,obs},\text{log}(g)_{\rm obs},\text{[Fe/H]}_{\rm obs}) \\ \times p(T_{\rm eff,\star,iso},t_{\rm \star, iso}|T_{\rm eff,\star,obs},\text{log}(g)_{\rm obs},\text{[Fe/H]}_{\rm obs}) dT_{\rm eff,\star,iso} dt_{\rm \star,iso} &, A_{\rm obs}=1.
    \end{cases}
\end{align}

We note that the $p(T_{\rm eff,\star,obs},\text{log}(g)_{\rm obs},\text{[Fe/H]}_{\rm obs})$ term, which shows up in the equations for each hypothesis, can be pulled out of the integrals and will cancel when odds ratios are calculated, so we do not need to specify it.

\subsection{Results}
To facilitate better comparison with previous analyses, we begin by using the stellar sample of \cite{triaud2011} with the TEPCat parameters as described above.  We perform integrations in Julia \citep{juliapaper} using the HCubature package (https://github.com/JuliaMath/HCubature.jl), which uses an adaptive GenzMalik algorithm \citep{GENZ1980295}.  We use this package because, while {\tt scipy.integrate.nquad} was sufficient for the calculations performed in the 2:1 resonances case, we need a faster algorithm for this case. We obtain the following odds ratios:
\begin{align}
    &\frac{p(H_{\rm nat})}{p(H_{\rm nur})} =1.6\times10^3\nonumber\\
    &\frac{p(H_{\rm nur})}{p(H_{\rm ch})} =1.4\nonumber\\
    &\frac{p(H_{\rm nat})}{p(H_{\rm ch})}=2.2\times10^3\nonumber.
\end{align}
These odds ratios mean we find very strong support for the Nature hypothesis (correlation due to temperature) over both Nurture (evolution over time) and Chance.  Regarding Nurture vs. Chance, there is nearly equal support for both hypotheses, which is the same result we obtained in Paper 1.  The odds ratios for Nature vs. Nurture and Nature vs. Chance are nearly an order of magnitude greater than those which we obtained in our previous analysis.  This result further strengthens our conclusion in support of \cite{winn2010}'s interpretation, that hot stars have high obliquities and cool stars have low obliquities.

In Paper 1, we used the same ages as \cite{triaud2011} did, whereas here, the ages are found via isochrones.  The magnitude of the difference between the original ages and the mean ages from the {\tt isochrones} program is less than 2 Gyr for all stars in the sample, and less than 1 Gyr for all but two stars.  Additionally, while we use the stellar temperatures from TEPCat in our original calculations as well as for the input to the {\tt isochrones} program, the mean temperatures output from {\tt isochrones} vary slightly from the input values; the difference is less than 35 K for all stars, and less than 15 K for all but four stars.  To check whether these differences have a significant effect on the outcome, we use the mean isochrone ages and mean isochrone temperatures to calculate the odds ratios without uncertainties.  We find results very close to those that we got in Paper 1: a Nature vs. Nurture ratio of 170, a Nurture vs. Chance ratio of 1.8, and a Nature vs. Chance ratio of 310.  To investigate how much of an effect the covariance of $T_{\rm eff,\star}$ and $t_\star$ has on the results, we marginalize over uncertainties in $T_{\rm eff,\star}$ and $t_\star$, as output by {\tt isochrones}, but treat $T_{\rm eff,\star}$ and $t_\star$ as independent of each other.  This yields odds ratios of Nature vs. Nurture of 840, Nurture vs. Chance of 1.8, and Nature vs. Chance of $1.5\times10^3$.  These results show that the increased support for the Nature hypothesis that we find here is primarily due to accounting for the uncertainties of $T_{\rm eff,\star}$ and $t_\star$, not from discrepancies between the ages used by \cite{triaud2011} and the isochrone-derived ages, and that the covariance between $T_{\rm eff,\star}$ and $t_\star$ plays a relatively minor role here.

\begin{figure}[ht]
    \includegraphics[width=3.5 in]{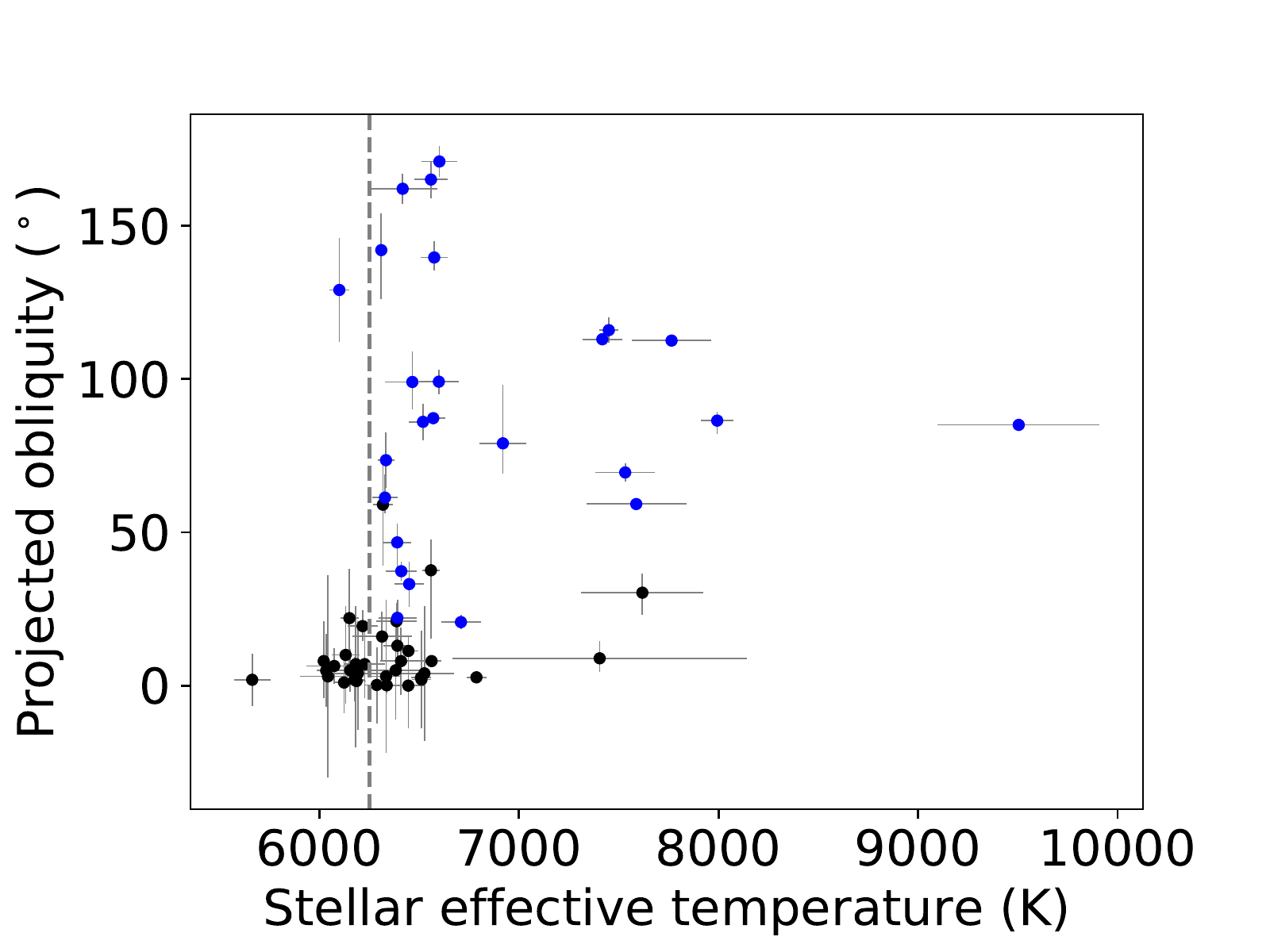}
    \includegraphics[width=3.5 in]{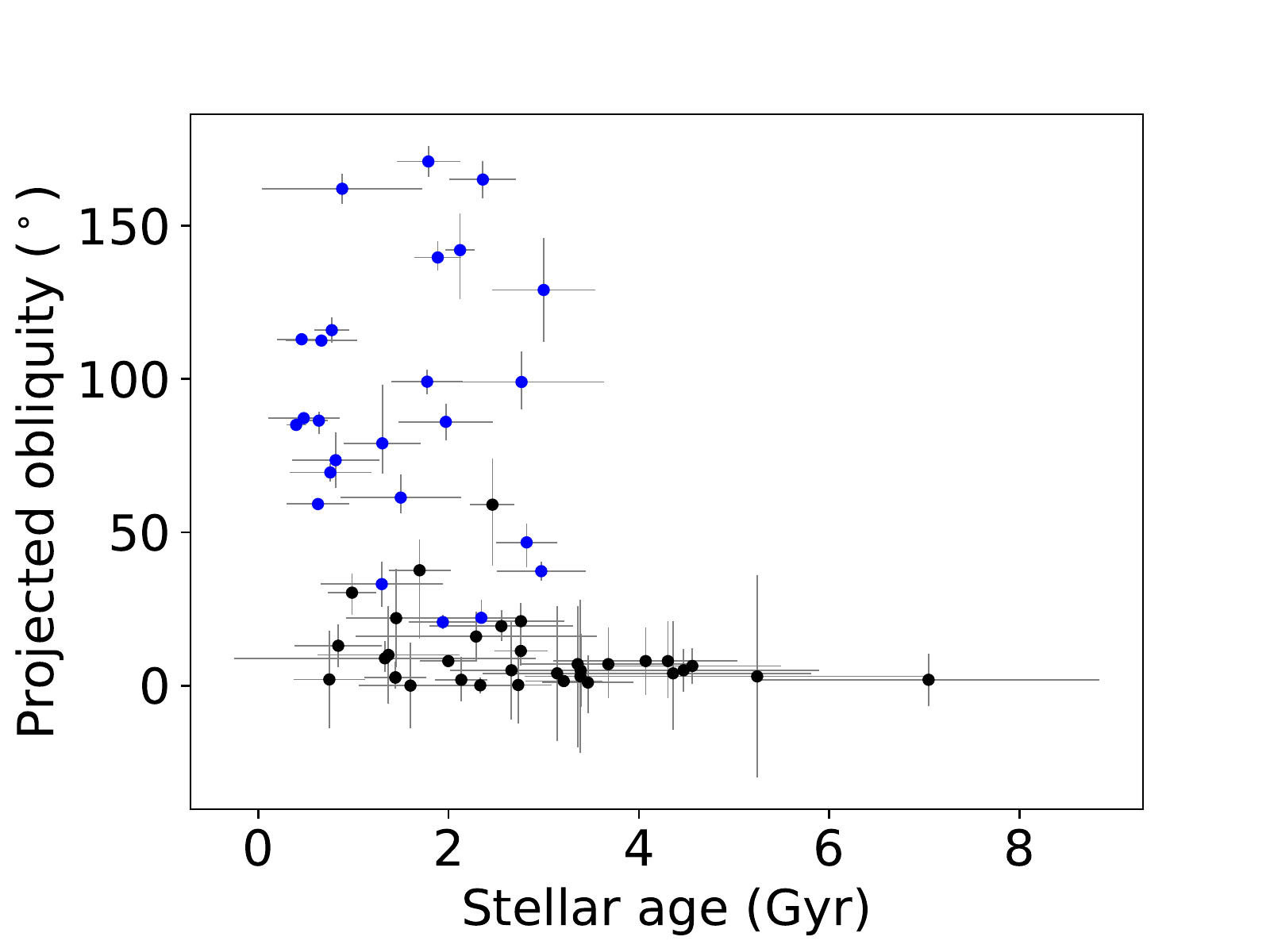}
    \caption{Projected obliquity ($\lambda$) versus stellar effective temperature (left) and mean age, in Gyr, from the {\tt isochrones} program (right).  Blue points represent systems that have $\lambda>10^\circ$ at the 3$\sigma$ level, and black points represent systems that do not.  The left panel also shows a dashed line at $T_{\rm eff,\star}=6250$ K, which we use to divide the sample into ``hot" and ``cool" stars.  Note that this sample only contains stars with $M_\star>1.2M_\odot$.}
    \label{fig:aligndataplots}
\end{figure}

We also test an updated sample of planetary systems with measured obliquities from Albrecht et al. (in prep).  This dataset consists of 59 stars with mass $M_\star>1.2M_\odot$, hosting planets with  $a/R_{\star} < 10$ and planet mass $M_{\rm P} > 0.5 M_{\rm J}$.  It contains 25 misaligned stars and 34 aligned stars.  We collected each star's $T_{\rm eff,\star,obs}$, log($g$)$_{\rm obs}$, and [Fe/H]$_{\rm obs}$ from the TEPCat catalog on 2021 June 16.  For two systems -- MASCARA-1 and MASCARA-4 -- the values for log($g$)$_{\rm obs}$ and [Fe/H]$_{\rm obs}$ are listed in the TEPCat catalog without any uncertainties; in these instances, we assign large uncertainties of 0.2 dex for log($g$)$_{\rm obs}$ and 0.1 dex for [Fe/H]$_{\rm obs}$.  The measured obliquities for this sample are plotted against the TEPCat stellar temperatures and mean ages from the {\tt isochrones} program in Figure \ref{fig:aligndataplots}.  This dataset yields the following odds ratios:
\begin{align}
    &\frac{p(H_{\rm nat})}{p(H_{\rm nur})} = 13\nonumber\\
    &\frac{p(H_{\rm nur})}{p(H_{\rm ch})} = 4.4\nonumber\\
    &\frac{p(H_{\rm nat})}{p(H_{\rm ch})}= 55\nonumber.
\end{align}

This updated sample shows a dramatic decrease in support for the Nature (temperature-driven) hypothesis compared to the original sample.  It appears that this shift is driven by the WASP-60 system, which has a star with $T_{\rm eff,\star}=6105 \pm 50$ K strongly misaligned at $\lambda=129\pm17^\circ$ -- a strong exception to the rule that cool stars have low obliquities.  When we redo the calculations with WASP-60 removed, we obtain odds ratios of:
\begin{align}
    &\frac{p(H_{\rm nat})}{p(H_{\rm nur})} = 3.7\times10^3\nonumber\\
    &\frac{p(H_{\rm nur})}{p(H_{\rm ch})} = 5.9\nonumber\\
    &\frac{p(H_{\rm nat})}{p(H_{\rm ch})}= 2.2\times10^4\nonumber.
\end{align}

However, WASP-60 is a borderline case, with a relatively large $a/R_\star$ of $8.5 \pm 0.4$, planet mass of 0.56$\pm 0.04$ Jupiter masses near our minimum, and a stellar effective temperature near the hot-cool boundary. Therefore WASP-60 may be a case of a system just out of reach of realignment. When we calculate the odds ratios using a cutoff of $a/R_\star<7$, which excludes WASP-60, we obtain $p(H_{\rm nat})/p(H_{\rm nur})=94$, $p(H_{\rm nur})/p(H_{\rm ch})=3.3$, and $p(H_{\rm nat})/p(H_{\rm ch})=310$, still strongly favoring Nature.  Assuming that it is acceptable for WASP-60 to be misaligned under the Nature hypothesis, our results of analyzing updated data indicate very strong support for the Nature hypothesis compared to Nurture or Chance.

Finally, it has been shown that stellar effective temperatures derived from different spectroscopic analysis pipelines can have large discrepancies, of 100 K or more \citep{furlan2018}.  Since the temperatures in the updated sample were not derived homogeneously, this means that our sample of hot stars may be contaminated by stars that are actually cool (and vice versa), potentially skewing the odds ratios.  We investigate this possible effect in two ways (excluding WASP-60 in each analysis).  First, we assume a systematic error in $T_{\rm eff,\star}$ of 100 K for each star, and add this in quadrature to the uncertainties from the TEPCat catalog before computing the isochrones. With these inflated uncertainties, we obtain odds ratios of $p(H_{\rm nat})/p(H_{\rm nur})=2.1\times10^3$, $p(H_{\rm nur})/p(H_{\rm ch})=6.2$, and $p(H_{\rm nat})/p(H_{\rm ch})=1.3\times10^4$.  This shows somewhat decreased, but still very strong, support for the Nature hypothesis over Nurture and Chance.  Second, we select a few of the hot stars near the 6250 K boundary to instead be cool.  We give each star with $6250\leq T_{\rm eff,\star} <6500$ a 50\% chance of having its temperature decreased by 250 K (putting it on the cool side), then recompute the isochrones and odds ratios.  This results in 11 hot stars switching to the cool side, including five misaligned stars.  The resulting probability for the Nature hypothesis is too small to compute.  This is not too surprising; the inclusion of the strong exception of WASP-60 is enough to make the Nature vs. Nurture odds ratio inconclusive, so adding several more strong exceptions would further exacerbate this effect.  We do not have any reason to suspect any particular hot star is actually cool, but this just underscores the need for accurate and homogeneously-derived stellar properties for large numbers of stars, for the proper assessment of perceived trends.

We conclude that the obliquities of stars with hot Jupiters are likely driven by stellar temperature rather than age.  Table \ref{tab:alignoddsratios} summarizes the odds ratios obtained without (from Paper 1) and with incorporating uncertainties, for the original samples as well as the updated sample, with and without WASP-60.

\begin{table}
	\centering
	\begin{tabular}{l|l|lll}
	    \multicolumn{5}{c}{STELLAR OBLIQUITIES} \\
		\hline
		& Without uncertainties& 
			    \multicolumn{3}{c}{With uncertainties} \\
		Ratio & \cite{triaud2011} (no unc.) & \cite{triaud2011} & Albrecht et al. & Albrecht et al., no WASP-60 \\
		\hline\hline
        $p(H_{\rm nat})/p(H_{\rm nur})$ & 210 & $1.6\times10^3$ & 13 & $3.7\times10^3$ \\
        $p(H_{\rm nur})/p(H_{\rm ch})$ & 1.4 & 1.4 & 4.4 & 5.9 \\
        $p(H_{\rm nat})/p(H_{\rm ch})$ & 310 & $2.2\times10^3$ & 55 & $2.2\times10^4$ \\
	    \hline
	\end{tabular}
	\caption{Odds ratios for the stellar obliquity case without (from Paper 1) and with incorporating measurement uncertainties, for the original sample as well as the updated sample.}
	\label{tab:alignoddsratios}
\end{table}

\section{Eccentricities of Hot Jupiters}
\label{section:eccentricities}
Multiple theories have been developed to explain the existence of hot Jupiters, and there is still some debate about which one plays the dominant role.  The lead contenders are high-eccentricity migration (e.g. \citealt{rasioford96}) and subsequent tidal circularization, disk migration \citep{goldreich1980,lin1986}, and in situ formation (e.g. \citealt{batygin2016,boley2016}).  Studying the eccentricities of hot Jupiters can offer a clue about which is the dominant mechanism because each will leave a different signature in the eccentricity distribution of hot Jupiters.  Disk migration and in situ formation are not expected to excite very large eccentricities (e.g. \citealt{petrovich14}), while high-eccentricity migration, as its name implies, sets planets on initially high-eccentricity orbits that circularize over time \citep{hotjupreview2018}.

\citet{quinn2014} found that hot Jupiters that were younger than their tidal circularization timescale $t_{\rm cir}$ tended to have eccentric orbits, while those older than $t_{\rm cir}$ tended to have circular orbits.  This is evidence that hot Jupiter orbits circularize over time, supporting high-eccentricity migration as the dominant formation path for this type of planet.  However, it was not obvious that separating their sample into groups older and younger than $t_{\rm cir}$ was the only effective way to divide the data into primarily circular and primarily eccentric planets.  In Paper 1 we showed that such a division could also effectively be achieved by separating the hot Jupiter sample into groups with small and large semimajor axes, with eccentric planets tending to be farther from their stars.  Astrophysically, this separation is plausible because eccentricities at larger semimajor axes are easier to excite and maintain (e.g. \citealt{petrovich14,duffell15}).  The strong dependence of $t_{\rm cir}$ on semimajor axis also means that a trend of eccentricity with semimajor axis may be seen, particularly in a sample with a small age range, if tidal circularization is at work.

In Paper 1, we compared evidence for the hypothesis that the distribution of hot Jupiter eccentricities is driven by age (Nurture) to the evidence that the eccentricity distribution is shaped by semimajor axis (Nature).  We found that the data very strongly supported a correlation due to age over a correlation due to semimajor axis, as well as over a Chance relation in which eccentricity is not at all related to other system properties.  Here, we build upon that result by using the modified framework to incorporate measurement uncertainties and examine how the uncertainties affect our results.

Rather than using the measured value of a planet's eccentricity, we only consider whether it means that a planet is on a circular or eccentric orbit.  We represent the eccentricity state with the binary parameter $E$, where $E=1$ if the planet is eccentric and $E=0$ if the planet is circular.  We assume that $E$ can only evolve from 1 to 0.  $E_{\rm obs}$ represents the observed eccentricity state of a system, and $E$ represents the true eccentricity state of a system.  We consider a planet to be eccentric if it has eccentricity $e>0$ at the 3$\sigma$ level.  We consider this criterion to mean that $E_{\rm obs}=E$, because it means that the eccentric group only contains systems with eccentricities measured well enough to be confidently classified as eccentric.  Accordingly, we will drop the $p(E_{\rm obs}|E)$ term from the following equations.

In the Nurture (age-dependent) hypothesis, we use the tidal circularization timescale $t_{\rm cir}$, which represents the decay rate of the orbital eccentricity and is given by \citep{socrates2012}:
\begin{equation}
\label{eqn:tcir}
    t_{\rm cir}=\frac{Q_{\rm P} M_{\rm P} a^{6.5}}{6\pi k_{\rm L} G^{0.5} M_{\star}^{1.5} R_{\rm P}^5}.
\end{equation}
Here, $M_{\rm P}$ is the planetary mass, $a$ is the semimajor axis, $M_\star$ is the stellar mass, and $R_{\rm P}$ is the planetary radius.  Following \citet{socrates2012}, we set $k_{\rm L}$, the Love number, to 0.38.

The planetary tidal quality factor $Q_{\rm P}$ is a rough measure of the efficiency of tidal dissipation in a planet.  It is difficult to constrain, as it likely depends on a variety of planetary parameters.  \cite{yoderpeale1981} constrained Jupiter's tidal quality factor to $6\times10^4 < Q_{\rm J} < 2\times10^6$.  For short period giant planets, it is generally assumed to be near $10^6$ (e.g. \citealt{ogilvielin2004,jackson2008}).  $Q_{\rm P}$ provides the primary source of uncertainty in the calculation of $t_{\rm cir}$, so in Paper 1 we parameterized the equation for the Nurture hypothesis in terms of $Q_{\rm P}$ instead of $t_{\rm cir}$.  We give $Q_{\rm P}$ a lognormal prior with mean $\mu$ and standard deviation $\sigma$, which are both hyperparameters.  We assign $\mu$ a uniform hyperprior U(2,8), and $\sigma$ a uniform hyperprior U(0,5), both in log$_{10}$ space.  These ranges are designed to cover the wide span and variety of possible values of $Q_{\rm P}$ among the hot Jupiter sample.  Then with accounting for measurement uncertainties, the equation for the Nurture hypothesis (derived from Eqn. \ref{eqn:gennurturebinary} and analogous to Eqn. 38 in Paper 1) is:
\begin{align}
    \label{eqn:eccnur}
    p(E_{\rm obs}&,t_{\rm \star,obs},a_{\rm obs},R_{\rm P,obs},M_{\rm P,obs},M_{\rm \star,obs}|f_0,\mu,\sigma)\nonumber\\ &=
    \begin{cases}
    \iiiint \int p(t_{\rm \star,obs}|t_\star) p(a_{\rm obs}|a) p(R_{\rm P,obs}|R_{\rm P}) p(M_{\rm P,obs}|M_{\rm P}) p(M_{\rm \star,obs}|M_\star) \\ \times p(t_{\star},a,R_{\rm P},M_{\rm P},M_{\star})\left[1-f_0+f_0\int_0^{Q_{\rm P,crit}} p(Q_{\rm P}|\mu,\sigma)dQ_{\rm P}\right] dt_\star da dR_{\rm P}dM_{\rm P}dM_\star &,E_{\rm obs}=0\\
    \iiiint \int p(t_{\rm \star,obs}|t_\star) p(a_{\rm obs}|a) p(R_{\rm P,obs}|R_{\rm P}) p(M_{\rm P,obs}|M_{\rm P}) p(M_{\rm \star,obs}|M_\star) \\ \times p(t_{\star},a,R_{\rm P},M_{\rm P},M_{\star})f_0\left[\int_{Q_{\rm P,crit}}^\infty p(Q_{\rm P}|\mu,\sigma)dQ_{\rm P}\right] dt_\star da dR_{\rm P}dM_{\rm P}dM_\star &,E_{\rm obs}=1.
    \end{cases}
\end{align}
Here, $Q_{\rm P,crit}$ is the value of $Q_{\rm P}$ for a given planet at which $t_{\rm cir}=t_\star$.  We use the hyperparameter $f_0$ to represent the fraction of planets that start out with eccentric orbits, and give it a uniform hyperprior from 0 to 1.  We have also assumed that the uncertainties of measured parameters are independent of each other.  For $t_\star$, $a$, $R_{\rm P}$, $M_{\rm P}$, and $M_\star$, we use uniform priors that encompass the observed values of these parameters in our sample.

The Nature hypothesis for the eccentricities case says that planets close to their stars have circular orbits and those farther out may be eccentric.  In Paper 1, we introduced the hyperparameter $a_{\rm cut}$, the semimajor axis cut-off within which planets are all circular.  Beyond $a_{\rm cut}$, a planet may be either circular or eccentric.  We use the hyperparameter $f_{\rm ecc,out}$ to describe the fraction of eccentric systems outside of $a_{\rm cut}$.  This leads to the following for $p(E|a,a_{\rm cut})$:
\begin{align}
    p(E|a< a_{\rm cut},f_{\rm ecc,out})=
    \begin{cases}
    1&,E=0\\
    0&,E=1
    \end{cases}
\end{align}
\begin{align}
    p(E|a\geq a_{\rm cut},f_{\rm ecc,out})=
    \begin{cases}
    1-f_{\rm ecc,out}&,E=0\\
    f_{\rm ecc,out}&,E=1.
    \end{cases}
\end{align}

We give $a_{\rm cut}$ a uniform hyperprior from 0 to 0.1 AU, and we give $f_{\rm ecc,out}$ a uniform hyperprior from 0 to 1. With uncertainties, which we again assume to be independent, the equation for the Nature hypothesis (derived from Eqn. \ref{eqn:gennature} and analogous to Eqn. 39 in Paper 1) is:
\begin{align}
    \label{eqn:eccnat}
    p(E_{\rm obs}&,t_{\rm \star,obs},a_{\rm obs},R_{\rm P,obs},M_{\rm P,obs},M_{\rm \star,obs}|a_{\rm cut},f_{\rm ecc,out})\nonumber\\ &= \iiiint \int p(t_{\rm \star,obs}|t_\star) p(a_{\rm obs}|a) p(R_{\rm P,obs}|R_{\rm P}) p(M_{\rm P,obs}|M_{\rm P}) p(M_{\rm \star,obs}|M_\star) \nonumber \\ &\times p(E_{\rm obs}|a,a_{\rm cut},f_{\rm ecc,out})p(t_{\star},a,R_{\rm P},M_{\rm P},M_{\star}) dt_\star da dR_{\rm P}dM_{\rm P}dM_\star.
\end{align}

In the Chance hypothesis, there is no relation between hot Jupiter eccentricities and any other system properties considered. The proportion of eccentric planets is only governed by the hyperparameter $f$, the overall fraction of planets that have eccentric orbits.  Then with uncertainties, which we assume to be independent, the equation for the Chance hypothesis (derived from Eqn. \ref{eqn:genchancebinary} and analogous to Eqn. 43 in Paper 1) is:
\begin{align}
    \label{eqn:eccchance}
    p(E_{\rm obs}&,t_{\rm \star,obs},a_{\rm obs},R_{\rm P,obs},M_{\rm P,obs},M_{\rm \star,obs}|f)\nonumber \\ &=
    \begin{cases}
    \iiiint \int p(t_{\rm \star,obs}|t_\star) p(a_{\rm obs}|a) p(R_{\rm P,obs}|R_{\rm P}) p(M_{\rm P,obs}|M_{\rm P}) p(M_{\rm \star,obs}|M_\star) \\ p(t_{\star},a,R_{\rm P},M_{\rm P},M_{\star})(1-f)dt_\star da dR_{\rm P}dM_{\rm P}dM_\star&,E_{\rm obs}=0\\
    \iiiint \int p(t_{\rm \star,obs}|t_\star) p(a_{\rm obs}|a) p(R_{\rm P,obs}|R_{\rm P}) p(M_{\rm P,obs}|M_{\rm P}) p(M_{\rm \star,obs}|M_\star) \\ p(t_{\star},a,R_{\rm P},M_{\rm P},M_{\star})fdt_\star da dR_{\rm P}dM_{\rm P}dM_\star&,E_{\rm obs}=1.
    \end{cases}
\end{align}
We give $f$ a uniform hyperprior from 0 to 1.

\subsection{Results}
\begin{figure}[ht]
    \includegraphics[width=3.5 in]{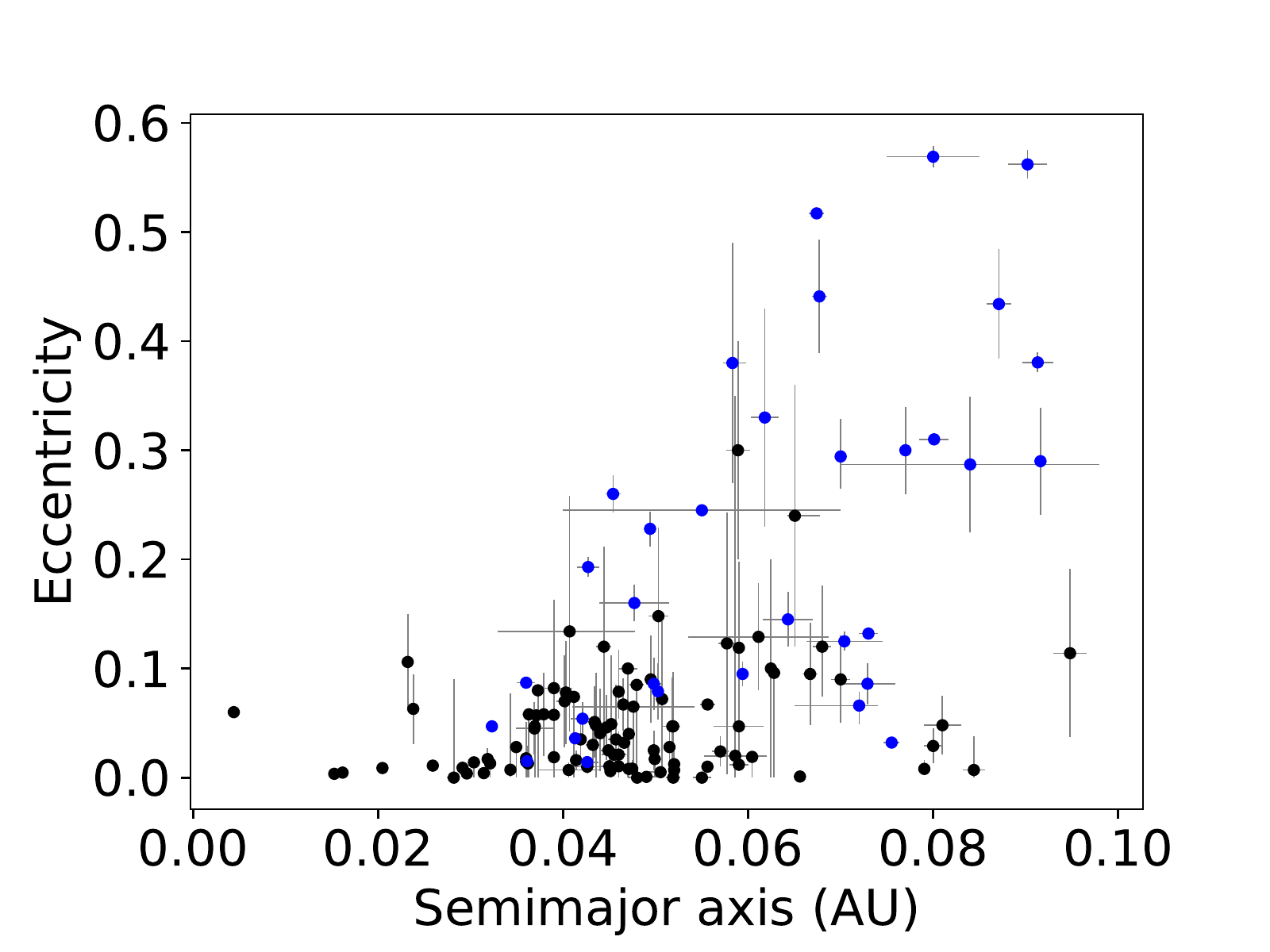}
    \includegraphics[width=3.5 in]{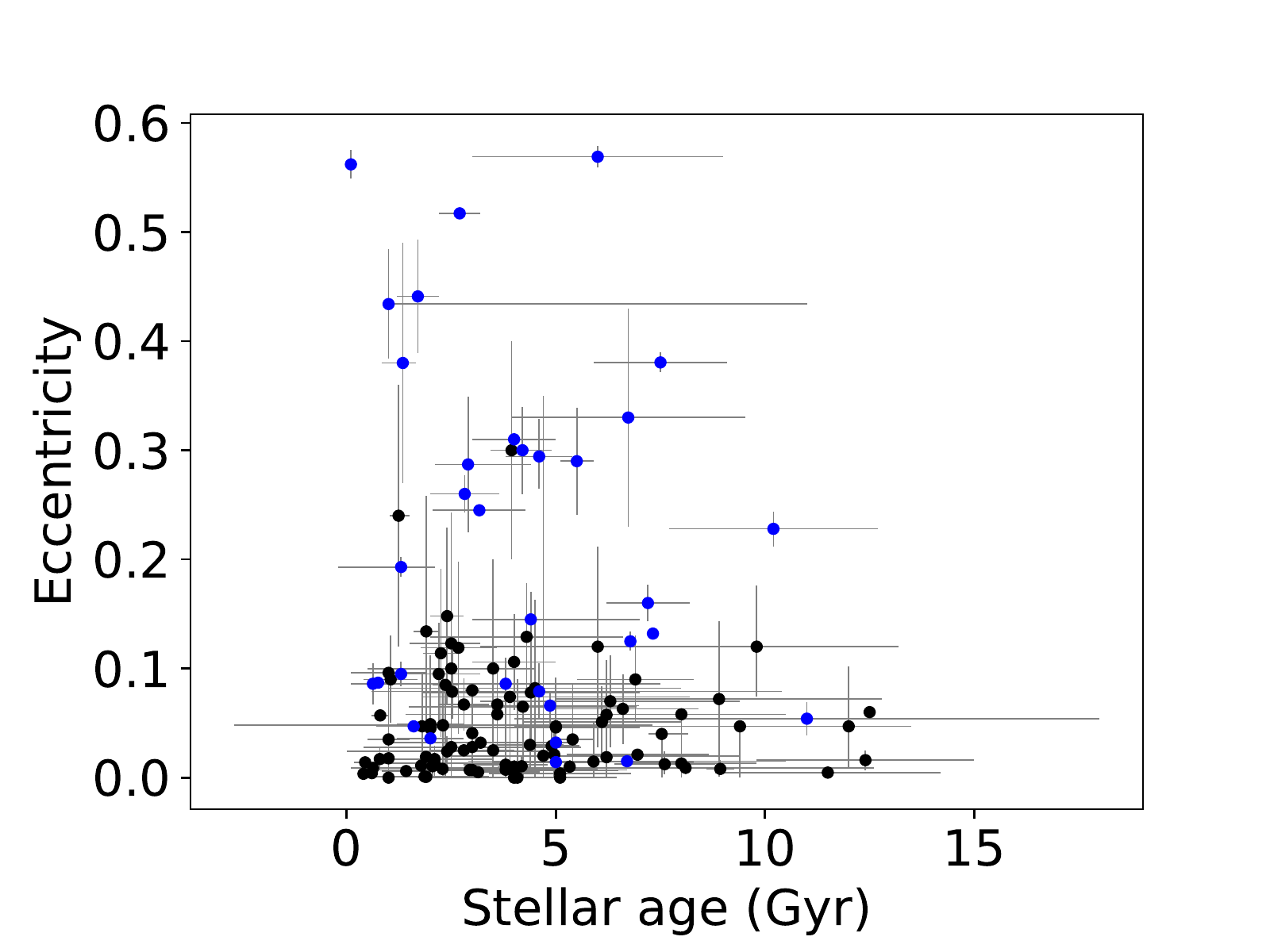}
    \includegraphics[width=3.5 in]{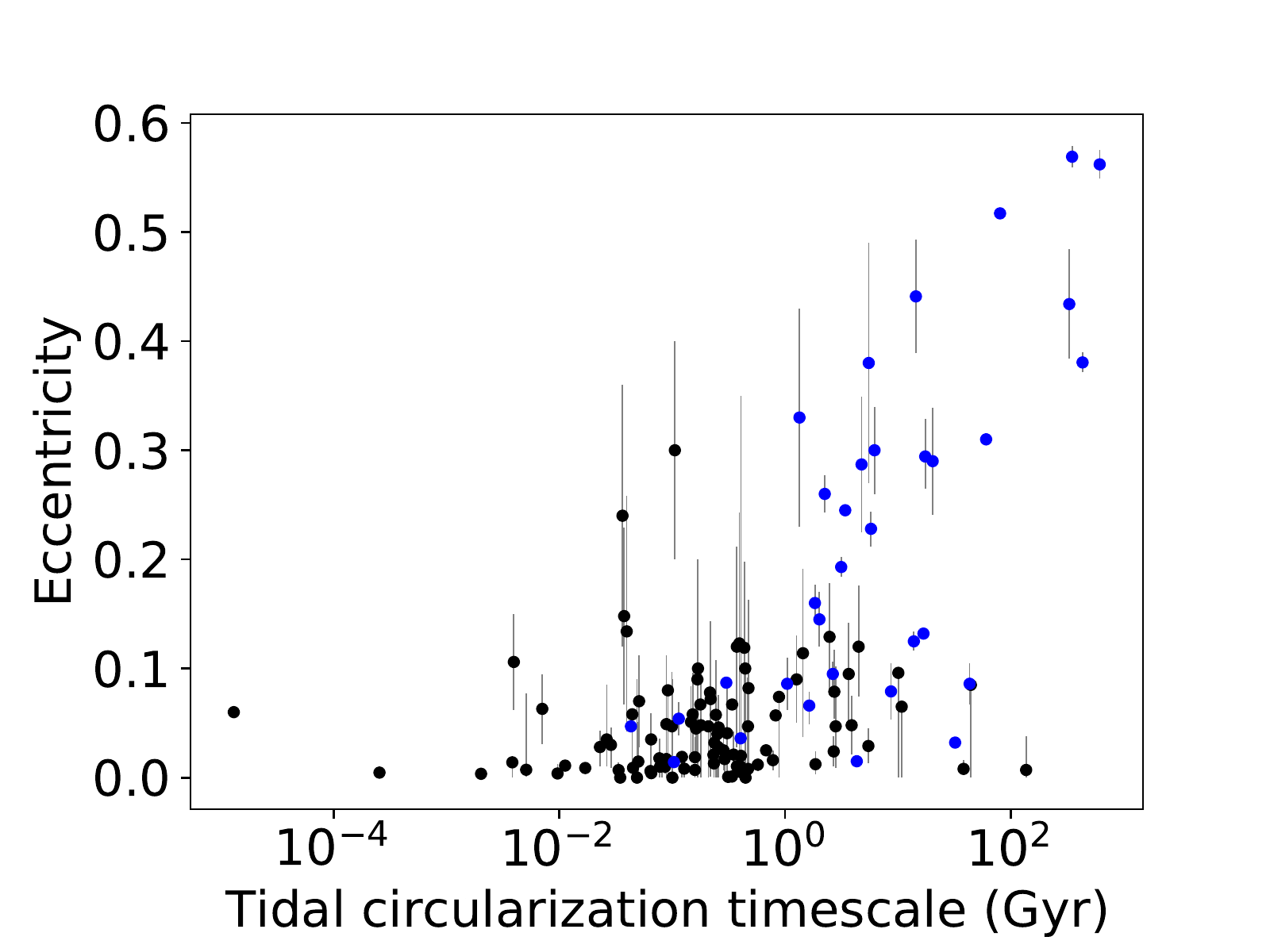}
    \caption{Eccentricity ($e$) versus semimajor axis (upper left), age in Gyr (upper right), and tidal circularization timescale in Gyr (lower left) for the sample of hot Jupiters we analyze in the eccentricities case.  Blue points represent systems that have $e>0$ at the 3$\sigma$ level, and black points represent systems that do not.}
    \label{fig:eccdataplots}
\end{figure}

For this case, we analyze the same sample that we used in Paper 1, and we refer the reader to Paper 1 for a link to a machine-readable table of the sample data.  This dataset consists of 130 stars hosting hot Jupiters (defined as having planetary mass $0.3 M_{\rm J}<M_{\rm P}<13M_{\rm J}$ and orbital period $P<10$ days) with measured stellar and planetary masses, semimajor axis, age, and eccentricity.  We obtained this data, as well as planetary radii for the planets which had them, from the Extrasolar Planets Encyclopaedia (exoplanet.eu; \citealt{schneider11}) on 2019 May 28.  For the 11 systems without reported planetary radii, we estimated the planetary radius using the relation in Eqn. 9 in \cite{weiss2013}.  The stellar ages have been derived in a variety of ways, mostly isochrones or evolutionary tracks.  Since this data was acquired fairly recently, we do not use an updated sample in this case.  Figure \ref{fig:eccdataplots} shows eccentricity versus semimajor axis, eccentricity versus stellar age, and eccentricity versus tidal circularization timescale for our sample.

We consider a planet to be eccentric ($E_{\rm obs}=1$) if it has $e>0$ with $3\sigma$ confidence, and circular ($E_{\rm obs}=0$) otherwise.  We have excluded planets with reported eccentricity of $e=0$ with no error bars.  If a planet has only an upper limit on the eccentricity, we classify it as circular if $e<0.1$ and exclude it otherwise.

For this case, we perform integrations in Julia using the HCubature package. As with the obliquities case, we choose this integrator over {\tt scipy.integrate.nquad} because of HCubature's greater speed.  With so many parameters to integrate over for so many systems, the full calculations of Eqns. \ref{eqn:eccnur}-\ref{eqn:eccchance} become prohibitively long.  Accordingly, we begin by integrating over just age, as it is typically the most uncertain of the parameter measurements, as well as the tidal quality factor $Q_{\rm P}$ (in the Nurture hypothesis) and the hyperparameters.  We treat the other observed parameters -- semimajor axis, planetary radius, planetary mass, and stellar mass -- as having no measurement uncertainty.  We obtain the following odds ratios:
\begin{align}
    &\frac{p(H_{\rm nur})}{p(H_{\rm nat})} =1.8\times10^8\nonumber\\
    &\frac{p(H_{\rm nur})}{p(H_{\rm ch})} =3.6\times10^8\nonumber\\
    &\frac{p(H_{\rm nat})}{p(H_{\rm ch})} =2.0\nonumber.
\end{align}
These results are nearly the same as those in Paper 1.  We noted in Paper 1 that the strong dependence of $t_{\rm cir}$ on $a$ and $R_{\rm P}$ means that uncertainties in those parameters could have a strong effect on the calculated $t_{\rm cir}$.  Integrating over uncertainties in both $a$ and $R_{\rm P}$ along with $t_\star$ ends up requiring too much computation time and memory.  Instead, we examined the effects of these parameters individually. Integrating over just uncertainties in $a$, treating all other observed parameters as having no measurement uncertainty, yields the following odds ratios:
\begin{align}
    &\frac{p(H_{\rm nur})}{p(H_{\rm nat})} =6.9\times10^7\nonumber\\
    &\frac{p(H_{\rm nur})}{p(H_{\rm ch})} =1.4\times10^8\nonumber\\
    &\frac{p(H_{\rm nat})}{p(H_{\rm ch})} =2.0\nonumber.
\end{align}
Finally, when integrating over just uncertainties in $R_{\rm P}$, we obtain the following odds ratios:
\begin{align}
    &\frac{p(H_{\rm nur})}{p(H_{\rm nat})} =1.1\times10^8\nonumber\\
    &\frac{p(H_{\rm nur})}{p(H_{\rm ch})} =2.2\times10^8\nonumber\\
    &\frac{p(H_{\rm nat})}{p(H_{\rm ch})} =2.0\nonumber.
\end{align}

Individually, the uncertainties on $t_\star$, $a$, and $R_{\rm P}$ do not make a substantial difference in the odds ratios, and so it does not seem likely that they would have a very profound influence when accounted for together.  The most significant difference comes from including $a$ uncertainties, which decreases support for the Nurture hypothesis by a factor of $\sim$2.  Table \ref{tab:eccoddsratios} summarizes the odds ratios obtained without (from Paper 1) and with incorporating uncertainties in $t_\star$, $a$, and $R_{\rm P}$.

\begin{table}
	\centering
	\begin{tabular}{l|l|lll}
	    \multicolumn{5}{c}{HOT JUPITER ECCENTRICITIES} \\
		\hline
		& Without uncertainties& 
			    \multicolumn{3}{c}{With uncertainties} \\
		Ratio & No unc. & $t_\star$ unc. & $a$ unc. & $R_{\rm P}$ unc. \\
		\hline\hline
        $p(H_{\rm nur})/p(H_{\rm nat})$ & $1.3\times10^8$ & $1.8\times10^8$ & $6.9\times10^7$ & $1.1\times10^8$ \\
        $p(H_{\rm nur})/p(H_{\rm ch})$ & $1.5\times10^8$ & $3.6\times10^8$ & $1.4\times10^8$ & $2.2\times10^8$ \\
        $p(H_{\rm nat})/p(H_{\rm ch})$ & 1.1 & 2.0 & 2.0 & 2.0 \\
	    \hline
	\end{tabular}
	\caption{Odds ratios for the eccentricities case without (from Paper 1) and with incorporating uncertainties in $t_\star$, $a$, and $R_{\rm P}$.}
	\label{tab:eccoddsratios}
\end{table}

Even when incorporating what we expect to be the most influential uncertainties, the evidence very strongly favors a trend of eccentricity driven by age, rather than by semimajor axis or random chance.  This result supports tidal circularization as a hot Jupiter formation mechanism.  We note that since we allow for an underlying circular population of hot Jupiters (represented by $1-f_0$), this does not mean that tidal circularization is the only formation mechanism for hot Jupiters, but does support that at least some hot Jupiters undergo this process.

\begin{figure}[ht]
    \includegraphics[width=3.5 in]{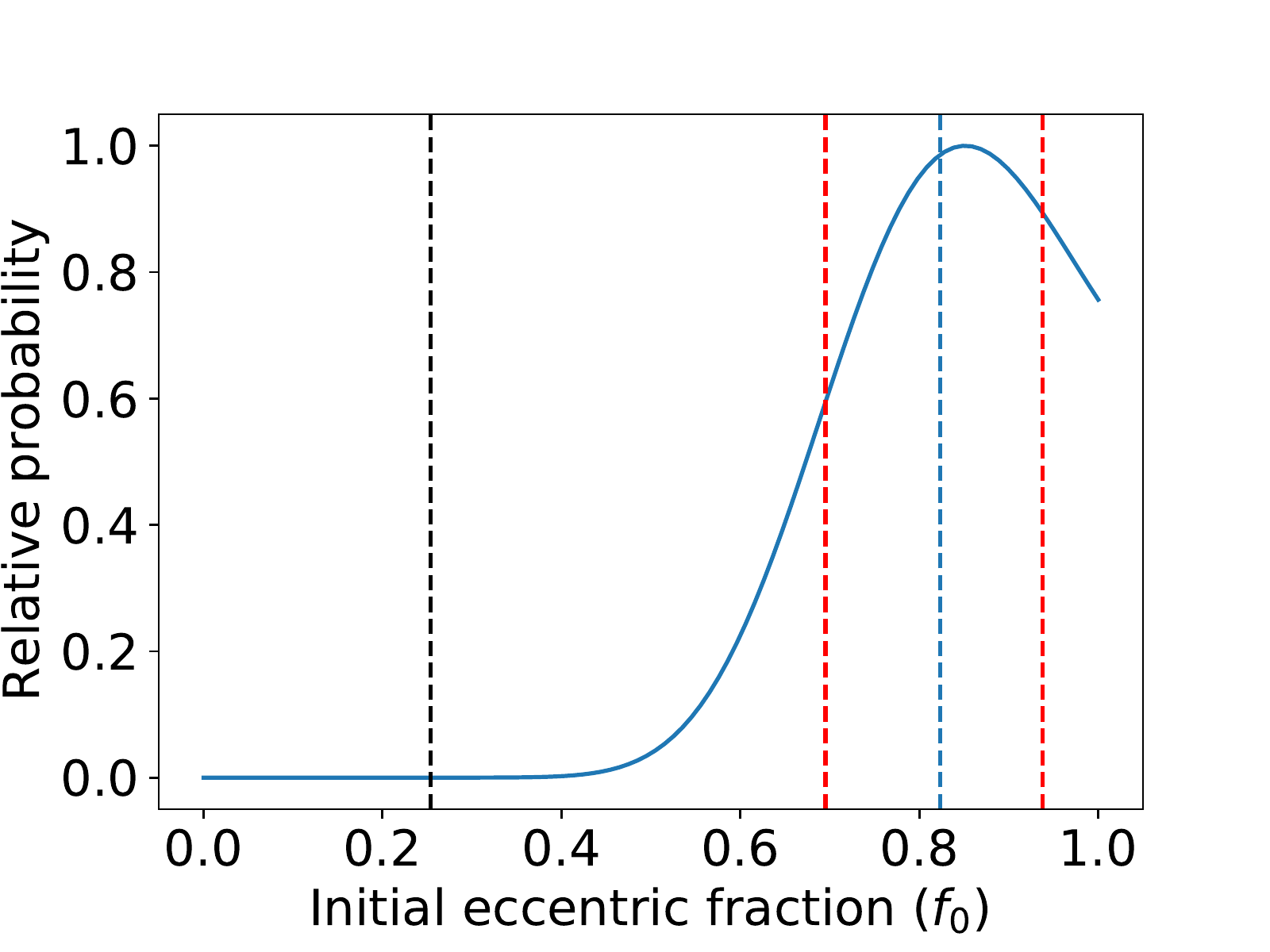}
    \caption{Posterior probability histogram of the initial eccentric fraction $f_0$, normalized such that the highest probability is 1.  The median is shown by the dashed blue line, and the red dashed lines mark the 68\% confidence interval.  For reference, the observed eccentric fraction is shown by the black dashed line; since the Nurture hypothesis involves evolution over time, we do not expect the observed eccentric fraction to match up with the median value for $f_0$.}
    \label{fig:f0hist}
\end{figure}

In Figure \ref{fig:f0hist} we display a posterior probability histogram of $f_0$, which has been marginalized over all other parameters and hyperparameters and normalized such that the highest probability is 1.  We draw $10^4$ random points from this distribution.  This sample has a median of 0.824, with a 68\% confidence interval of 0.692 to 0.935.  We thus conclude that under the Nurture hypothesis, at least half of the planets in our sample have undergone or are currently experiencing tidal circularization.

\section{Conclusion}
\label{section:conclusion}
We improve upon the Bayesian framework outlined in Paper 1 by formally incorporating measurement uncertainties (Section \ref{section:generalframework}).  We then reanalyze the data we used in Paper 1 with the updated framework and find that marginalizing over measurement uncertainties does not change our original conclusions.  We also analyze updated samples for both the 2:1 resonances and the obliquities cases, and find our previous conclusions still hold.  These results lead us to conclude that hot Jupiters are circularized over time, supporting high-eccentricity migration and tidal circularization (Section \ref{section:eccentricities}), and that the stellar obliquities of stars with hot Jupiters are driven by stellar effective temperature (Section \ref{section:obliquities}).  It remains unclear whether 2:1 orbital resonances are disrupted over time (Section \ref{section:resonances}).  Through the analysis of simulated data, we show that significant uncertainties, comparable to the uncertainties in the real data, in the measured stellar ages can, in some cases, obscure a true trend of 2:1 resonances with age.  However, an increased sample size will likely have a more significant effect on the odds ratio than just more precise ages.  Even with a large sample size, though, it will be difficult to confirm a trend of 2:1 resonances with age if the resonances get disrupted on a wide range of timescales.

While it is relatively simple to include uncertainties in the general equations as well as equations for specific applications, performing the complete calculations may, in some cases, present a prohibitive computational challenge.  Each observed parameter included in the equations means another integral to perform, and the computational time and memory required to do them all can pile up to impracticable levels.  Additionally, in Paper 1 we showed that in certain cases, some terms in the equations canceled out when the odds ratios were taken, eliminating the need to calculate those terms.  When uncertainties are fully incorporated, such cancellation may not be possible.  For example, in the obliquities case (Section \ref{section:obliquities}), we integrate over a joint posterior of stellar age and stellar effective temperature, $p(T_{\rm eff,\star,iso},t_{\rm \star, iso}|T_{\rm eff,\star,obs},\text{log}(g)_{\rm obs},\text{[Fe/H]}_{\rm obs})$. For an individual star, this term is, of course, the same under each hypothesis.  However, since age and temperature appear elsewhere in the equations, the integral over the joint posterior cannot be separated out and canceled when ratios are taken (specifically, stellar age appears in the integration limits over the alignment timescale $t_{\rm a}$ in the Nurture hypothesis, and the stellar effective temperature appears in the $p(A|T_{\rm eff,\star,iso})$ term, which describes how the alignment state $A$ depends on effective temperature, in the Nature hypothesis).  Without incorporating uncertainties, there is still a $p(T_{\rm eff,\star},t_\star)$ term, but no integral over $T_{\rm eff,\star}$ or $t_\star$, so $p(T_{\rm eff,\star},t_\star)$ would cancel in the odds ratios.  For applications of this framework with many variables to account for and/or a large sample size, the use of the simpler framework in Paper 1 may be preferable, accompanied by additional tests (bootstrapping, removing outliers, etc.) to explore the effect of highly uncertain measurements on the results.  Alternatively, one could choose to only account for uncertainties on parameters with large measurement errors, as we do for the eccentricities case here (see Section \ref{section:eccentricities}).

\acknowledgments We thank the referee for the helpful comments on this paper.  We thank Angie Wolfgang for helpful discussions.  We gratefully acknowledge support from NASA XRP 80NSSC18K0355.  This material is based upon work supported by the National Aeronautics and Space Administration under Grant No. 80NSSC20M0097 issued through the PA Space Grant Consortium.  The Center for Exoplanets and Habitable Worlds is supported by the Pennsylvania State University, the Eberly College of Science, and the Pennsylvania Space Grant Consortium.  Computations for this research were performed on the Pennsylvania State University’s Institute for Computational and Data Sciences’ Roar supercomputer.  This content is solely the responsibility of the authors and does not necessarily represent the views of the Institute for Computational and Data Sciences. We thank Andrew Polasky and Weinan Chen for input on parallel processing techniques. This research has made use of the NASA Exoplanet Archive, which is operated by the California Institute of Technology, under contract with the National Aeronautics and Space Administration under the Exoplanet Exploration Program.  This research has made use of data obtained from or tools provided by the portal exoplanet.eu of The Extrasolar Planets Encyclopaedia.  We acknowledge the use of the software packages NumPy \citep{numpy} in performing calculations and generating random numbers; Pandas \citep{pandas2010,pandas2011} in reading in data files; Matplotlib \citep{matplotlib} in generating plots; and Cython \citep{cython} and PyJulia \citep{pyjulia} in creating code that runs faster than pure Python.

\software{SciPy \citep{scipy},
NumPy \citep{numpy},
Matplotlib \citep{matplotlib},
Cython \citep{cython},
Pandas \citep{pandas2010,pandas2011},
PyJulia \citep{pyjulia},
Julia \citep{juliapaper},
isochrones \citep{isochrones}}

\bibliography{bibliography} \bibliographystyle{aasjournal}

\begin{thebibliography}{}
\expandafter\ifx\csname natexlab\endcsname\relax\def\natexlab#1{#1}\fi
\providecommand{\url}[1]{\href{#1}{#1}}
\providecommand{\dodoi}[1]{doi:~\href{http://doi.org/#1}{\nolinkurl{#1}}}
\providecommand{\doeprint}[1]{\href{http://ascl.net/#1}{\nolinkurl{http://ascl.net/#1}}}
\providecommand{\doarXiv}[1]{\href{https://arxiv.org/abs/#1}{\nolinkurl{https://arxiv.org/abs/#1}}}

\bibitem[{{Albrecht} {et~al.}(2012){Albrecht}, {Winn}, {Johnson}, {Howard},
  {Marcy}, {Butler}, {Arriagada}, {Crane}, {Shectman}, {Thompson}, {Hirano},
  {Bakos}, \& {Hartman}}]{albrecht2012}
{Albrecht}, S., {Winn}, J.~N., {Johnson}, J.~A., {et~al.} 2012, \apj, 757, 18,
  \dodoi{10.1088/0004-637X/757/1/18}

\bibitem[{{Arakaki} {et~al.}(2020){Arakaki}, {Bolewski}, {Deits}, {Fischer},
  {Johnson}, {Bussonnier}, {Norton}, {Haraldsson}, {Rocklin}, {Tsur}, {Shah},
  {Soto}, {Eslgastal}, {Kuthe}, {Jakirkham}, {Millea}, {Grahamgill},
  {Fnmdx111}, {Arslan}, {Lukas}, {Nadlinger}, {Besson}, {Olver}, {Zhao}, \&
  {Scls19fr}}]{pyjulia}
{Arakaki}, T., {Bolewski}, J., {Deits}, R., {et~al.} 2020, {JuliaPy/pyjulia:
  PyJulia v0.5.6}, v0.5.6,  Zenodo, \dodoi{10.5281/zenodo.4294939}

\bibitem[{{Batygin} {et~al.}(2016){Batygin}, {Bodenheimer}, \&
  {Laughlin}}]{batygin2016}
{Batygin}, K., {Bodenheimer}, P.~H., \& {Laughlin}, G.~P. 2016, \apj, 829, 114,
  \dodoi{10.3847/0004-637X/829/2/114}

\bibitem[{Behnel {et~al.}(2011)Behnel, Bradshaw, Citro, Dalcin, Seljebotn, \&
  Smith}]{cython}
Behnel, S., Bradshaw, R., Citro, C., {et~al.} 2011, Computing in Science \&
  Engineering, 13, 31

\bibitem[{{Berger} {et~al.}(2018){Berger}, {Huber}, {Gaidos}, \& {van
  Saders}}]{berger2018}
{Berger}, T.~A., {Huber}, D., {Gaidos}, E., \& {van Saders}, J.~L. 2018, \apj,
  866, 99, \dodoi{10.3847/1538-4357/aada83}

\bibitem[{{Berger} {et~al.}(2020{\natexlab{a}}){Berger}, {Huber}, {Gaidos},
  {van Saders}, \& {Weiss}}]{berger2020planets}
{Berger}, T.~A., {Huber}, D., {Gaidos}, E., {van Saders}, J.~L., \& {Weiss},
  L.~M. 2020{\natexlab{a}}, \aj, 160, 108, \dodoi{10.3847/1538-3881/aba18a}

\bibitem[{{Berger} {et~al.}(2020{\natexlab{b}}){Berger}, {Huber}, {van Saders},
  {Gaidos}, {Tayar}, \& {Kraus}}]{berger2020}
{Berger}, T.~A., {Huber}, D., {van Saders}, J.~L., {et~al.} 2020{\natexlab{b}},
  \aj, 159, 280, \dodoi{10.3847/1538-3881/159/6/280}

\bibitem[{Bezanson {et~al.}(2017)Bezanson, Edelman, Karpinski, \&
  Shah}]{juliapaper}
Bezanson, J., Edelman, A., Karpinski, S., \& Shah, V.~B. 2017, SIAM review, 59,
  65.
\newblock \url{https://doi.org/10.1137/141000671}

\bibitem[{{Boley} {et~al.}(2016){Boley}, {Granados Contreras}, \&
  {Gladman}}]{boley2016}
{Boley}, A.~C., {Granados Contreras}, A.~P., \& {Gladman}, B. 2016, \apjl, 817,
  L17, \dodoi{10.3847/2041-8205/817/2/L17}

\bibitem[{{Christiansen} {et~al.}(2019){Christiansen}, {Beichman}, {Ciardi}, \&
  {Huber}}]{christiansen2019}
{Christiansen}, J., {Beichman}, C., {Ciardi}, D.~R., \& {Huber}, D. 2019,
  \baas, 51, 312.
\newblock \doarXiv{1903.09110}

\bibitem[{{Dawson} \& {Johnson}(2018)}]{hotjupreview2018}
{Dawson}, R.~I., \& {Johnson}, J.~A. 2018, \araa, 56, 175,
  \dodoi{10.1146/annurev-astro-081817-051853}

\bibitem[{{Dong} \& {Dawson}(2016)}]{dd2016}
{Dong}, R., \& {Dawson}, R. 2016, \apj, 825, 77,
  \dodoi{10.3847/0004-637X/825/1/77}

\bibitem[{{Duffell} \& {Chiang}(2015)}]{duffell15}
{Duffell}, P.~C., \& {Chiang}, E. 2015, \apj, 812, 94,
  \dodoi{10.1088/0004-637X/812/2/94}

\bibitem[{{Furlan} {et~al.}(2018){Furlan}, {Ciardi}, {Cochran}, {Everett},
  {Latham}, {Marcy}, {Buchhave}, {Endl}, {Isaacson}, {Petigura}, {Gautier},
  {Huber}, {Bieryla}, {Borucki}, {Brugamyer}, {Caldwell}, {Cochran}, {Howard},
  {Howell}, {Johnson}, {MacQueen}, {Quinn}, {Robertson}, {Mathur}, \&
  {Batalha}}]{furlan2018}
{Furlan}, E., {Ciardi}, D.~R., {Cochran}, W.~D., {et~al.} 2018, \apj, 861, 149,
  \dodoi{10.3847/1538-4357/aaca34}

\bibitem[{{Gaia Collaboration} {et~al.}(2016){Gaia Collaboration}, {Prusti},
  {de Bruijne}, {Brown}, {Vallenari}, {Babusiaux}, {Bailer-Jones}, {Bastian},
  {Biermann}, {Evans}, \& et~al.}]{gaiamission}
{Gaia Collaboration}, {Prusti}, T., {de Bruijne}, J.~H.~J., {et~al.} 2016,
  \aap, 595, A1, \dodoi{10.1051/0004-6361/201629272}

\bibitem[{Genz \& Malik(1980)}]{GENZ1980295}
Genz, A., \& Malik, A. 1980, Journal of Computational and Applied Mathematics,
  6, 295, \dodoi{https://doi.org/10.1016/0771-050X(80)90039-X}

\bibitem[{{Goldreich} \& {Tremaine}(1980)}]{goldreich1980}
{Goldreich}, P., \& {Tremaine}, S. 1980, \apj, 241, 425, \dodoi{10.1086/158356}

\bibitem[{Harris {et~al.}(2020)Harris, Millman, van~der Walt, Gommers,
  Virtanen, Cournapeau, Wieser, Taylor, Berg, Smith, Kern, Picus, Hoyer, van
  Kerkwijk, Brett, Haldane, del R{'{\i}}o, Wiebe, Peterson,
  G{'{e}}rard-Marchant, Sheppard, Reddy, Weckesser, Abbasi, Gohlke, \&
  Oliphant}]{numpy}
Harris, C.~R., Millman, K.~J., van~der Walt, S.~J., {et~al.} 2020, Nature, 585,
  357, \dodoi{10.1038/s41586-020-2649-2}

\bibitem[{Hunter(2007)}]{matplotlib}
Hunter, J.~D. 2007, Computing In Science \& Engineering, 9, 90

\bibitem[{{Izidoro} {et~al.}(2017){Izidoro}, {Ogihara}, {Raymond},
  {Morbidelli}, {Pierens}, {Bitsch}, {Cossou}, \& {Hersant}}]{izid17}
{Izidoro}, A., {Ogihara}, M., {Raymond}, S.~N., {et~al.} 2017, \mnras, 470,
  1750, \dodoi{10.1093/mnras/stx1232}

\bibitem[{{Jackson} {et~al.}(2008){Jackson}, {Greenberg}, \&
  {Barnes}}]{jackson2008}
{Jackson}, B., {Greenberg}, R., \& {Barnes}, R. 2008, \apj, 678, 1396,
  \dodoi{10.1086/529187}

\bibitem[{Jeffreys(1961)}]{jeffreys1961}
Jeffreys, H. 1961, Theory of probability, 3rd edn. (Oxford: Clarendon Press)

\bibitem[{Kass \& Raftery(1995)}]{kass1995}
Kass, R.~E., \& Raftery, A.~E. 1995, Journal of the American Statistical
  Association, 90, 773.
\newblock \url{http://www.jstor.org/stable/2291091}

\bibitem[{{Koriski} \& {Zucker}(2011)}]{kz2011}
{Koriski}, S., \& {Zucker}, S. 2011, \apjl, 741, L23,
  \dodoi{10.1088/2041-8205/741/1/L23}

\bibitem[{{Kraft}(1967)}]{kraft1967}
{Kraft}, R.~P. 1967, \apj, 150, 551, \dodoi{10.1086/149359}

\bibitem[{{Lin} \& {Papaloizou}(1986)}]{lin1986}
{Lin}, D.~N.~C., \& {Papaloizou}, J. 1986, \apj, 309, 846,
  \dodoi{10.1086/164653}

\bibitem[{{Mamajek} \& {Hillenbrand}(2008)}]{mamhill2008}
{Mamajek}, E.~E., \& {Hillenbrand}, L.~A. 2008, \apj, 687, 1264,
  \dodoi{10.1086/591785}

\bibitem[{McKinney(2010)}]{pandas2010}
McKinney, W. 2010, in Proceedings of the 9th Python in Science Conference, Vol.
  445, Austin, TX, 51--56

\bibitem[{McKinney(2011)}]{pandas2011}
McKinney, W. 2011, Python for High Performance and Scientific Computing, 14

\bibitem[{{Morton}(2015)}]{isochrones}
{Morton}, T.~D. 2015, {isochrones: Stellar model grid package}.
\newblock \doeprint{1503.010}

\bibitem[{{NASA Exoplanet Archive}(2021{\natexlab{a}})}]{psrv}
{NASA Exoplanet Archive}. 2021{\natexlab{a}}, Planetary Systems, Version:
  2021-08-26 12:07,  NExScI-Caltech/IPAC, \dodoi{10.26133/NEA12}

\bibitem[{{NASA Exoplanet Archive}(2021{\natexlab{b}})}]{pskepler}
---. 2021{\natexlab{b}}, Planetary Systems, Version: 2021-11-12 13:11,
  NExScI-Caltech/IPAC, \dodoi{10.26133/NEA12}

\bibitem[{{Ogilvie} \& {Lin}(2004)}]{ogilvielin2004}
{Ogilvie}, G.~I., \& {Lin}, D.~N.~C. 2004, \apj, 610, 477,
  \dodoi{10.1086/421454}

\bibitem[{{Petrovich} {et~al.}(2014){Petrovich}, {Tremaine}, \&
  {Rafikov}}]{petrovich14}
{Petrovich}, C., {Tremaine}, S., \& {Rafikov}, R. 2014, \apj, 786, 101,
  \dodoi{10.1088/0004-637X/786/2/101}

\bibitem[{{Quinn} {et~al.}(2014){Quinn}, {White}, {Latham}, {Buchhave},
  {Torres}, {Stefanik}, {Berlind}, {Bieryla}, {Calkins}, {Esquerdo}, {F{\H
  u}r{\'e}sz}, {Geary}, \& {Szentgyorgyi}}]{quinn2014}
{Quinn}, S.~N., {White}, R.~J., {Latham}, D.~W., {et~al.} 2014, \apj, 787, 27,
  \dodoi{10.1088/0004-637X/787/1/27}

\bibitem[{{Rasio} \& {Ford}(1996)}]{rasioford96}
{Rasio}, F.~A., \& {Ford}, E.~B. 1996, Science, 274, 954,
  \dodoi{10.1126/science.274.5289.954}

\bibitem[{{Safsten} {et~al.}(2020){Safsten}, {Dawson}, \&
  {Wolfgang}}]{safstenetal2020}
{Safsten}, E.~D., {Dawson}, R.~I., \& {Wolfgang}, A. 2020, \aj, 160, 214,
  \dodoi{10.3847/1538-3881/abb536}

\bibitem[{{Schlaufman}(2010)}]{schlaufman2010}
{Schlaufman}, K.~C. 2010, \apj, 719, 602, \dodoi{10.1088/0004-637X/719/1/602}

\bibitem[{{Schneider} {et~al.}(2011){Schneider}, {Dedieu}, {Le Sidaner},
  {Savalle}, \& {Zolotukhin}}]{schneider11}
{Schneider}, J., {Dedieu}, C., {Le Sidaner}, P., {Savalle}, R., \&
  {Zolotukhin}, I. 2011, \aap, 532, A79, \dodoi{10.1051/0004-6361/201116713}

\bibitem[{{Socrates} {et~al.}(2012){Socrates}, {Katz}, \&
  {Dong}}]{socrates2012}
{Socrates}, A., {Katz}, B., \& {Dong}, S. 2012, arXiv e-prints,
  arXiv:1209.5724.
\newblock \doarXiv{1209.5724}

\bibitem[{{Soderblom}(2010)}]{soderblom2010}
{Soderblom}, D.~R. 2010, \araa, 48, 581,
  \dodoi{10.1146/annurev-astro-081309-130806}

\bibitem[{{Southworth}(2011)}]{tepcat}
{Southworth}, J. 2011, \mnras, 417, 2166,
  \dodoi{10.1111/j.1365-2966.2011.19399.x}

\bibitem[{{Thommes} {et~al.}(2008){Thommes}, {Bryden}, {Wu}, \&
  {Rasio}}]{thommes2008}
{Thommes}, E.~W., {Bryden}, G., {Wu}, Y., \& {Rasio}, F.~A. 2008, \apj, 675,
  1538, \dodoi{10.1086/525244}

\bibitem[{{Triaud}(2011)}]{triaud2011}
{Triaud}, A.~H.~M.~J. 2011, \aap, 534, L6, \dodoi{10.1051/0004-6361/201117713}

\bibitem[{{Virtanen} {et~al.}(2020){Virtanen}, {Gommers}, {Oliphant},
  {Haberland}, {Reddy}, {Cournapeau}, {Burovski}, {Peterson}, {Weckesser},
  {Bright}, {van der Walt}, {Brett}, {Wilson}, {Jarrod Millman}, {Mayorov},
  {Nelson}, {Jones}, {Kern}, {Larson}, {Carey}, {Polat}, {Feng}, {Moore}, {Vand
  erPlas}, {Laxalde}, {Perktold}, {Cimrman}, {Henriksen}, {Quintero}, {Harris},
  {Archibald}, {Ribeiro}, {Pedregosa}, {van Mulbregt}, \&
  {Contributors}}]{scipy}
{Virtanen}, P., {Gommers}, R., {Oliphant}, T.~E., {et~al.} 2020, Nature
  Methods, 17, 261, \dodoi{https://doi.org/10.1038/s41592-019-0686-2}

\bibitem[{{Weiss} {et~al.}(2013){Weiss}, {Marcy}, {Rowe}, {Howard}, {Isaacson},
  {Fortney}, {Miller}, {Demory}, {Fischer}, {Adams}, {Dupree}, {Howell},
  {Kolbl}, {Johnson}, {Horch}, {Everett}, {Fabrycky}, \& {Seager}}]{weiss2013}
{Weiss}, L.~M., {Marcy}, G.~W., {Rowe}, J.~F., {et~al.} 2013, \apj, 768, 14,
  \dodoi{10.1088/0004-637X/768/1/14}

\bibitem[{{Winn} {et~al.}(2010){Winn}, {Fabrycky}, {Albrecht}, \&
  {Johnson}}]{winn2010}
{Winn}, J.~N., {Fabrycky}, D., {Albrecht}, S., \& {Johnson}, J.~A. 2010, \apjl,
  718, L145, \dodoi{10.1088/2041-8205/718/2/L145}

\bibitem[{{Yoder} \& {Peale}(1981)}]{yoderpeale1981}
{Yoder}, C.~F., \& {Peale}, S.~J. 1981, \icarus, 47, 1,
  \dodoi{10.1016/0019-1035(81)90088-9}

\end{thebibliography}

\appendices
\section{Accounting for Uncertainty in Period Ratio in 2:1 Resonances case}
\label{sec:appendix}
When studying the case of 2:1 orbital resonances in Section \ref{section:resonances}, we consider the 2:1 resonant state of a system $R$, rather than directly using the period ratio or normalized commensurability proximity (NCP; Eqn. \ref{eqn:ncp}) parameters themselves. The resulting $p(R_{\rm obs}|R)$ term in the equations represents the probability of seeing the observed resonance state given the true resonance state of a system.  Since the majority of the systems in our samples are more than 3$\sigma$ away from the 2:1 resonance threshold of $\delta=0.1$ -- i.e. are confidently either near a 2:1 resonance or not -- based on period ratio uncertainties, we assumed $R_{\rm obs}=R$ and dropped the $p(R_{\rm obs}|R)$ term from the equations.  Here, we fully incorporate uncertainties in period ratio to address the few systems in our samples whose period ratio errors put them within 3$\sigma$ of the 2:1 resonance threshold.  In the RV sample, two systems are within 3$\sigma$, two systems are within 2$\sigma$, and one system is within 1$\sigma$.  In both of the \textit{Kepler} samples, no systems are within 3$\sigma$.

It is not straightforward to translate uncertainty in period ratio to a formulation for $p(R_{\rm obs}|R)$.  Instead, we rewrite the equations in terms of the period ratio itself, $r$, and the observed value of the period ratio, $r_{\rm obs}$.  This formulation will allow us to marginalize over $r$.  A certain range of values of $r$, surrounding $r=2$, yield NCP values of $\delta<0.1$ and thus are considered to be near a 2:1 resonance.  We represent the lower and upper bounds of the range as $r_{\rm l}$ and $r_{\rm h}$, respectively.  Within the range of $r_{\rm l}$-$r_{\rm h}$, the rest of the equation takes the form for $R=1$; otherwise, it takes the form for $R=0$.  Then the likelihood equations become
\begin{align}
\label{eqn:rnur}
  p(r_{\rm obs},t_{\rm \star,obs}|\mu & ,\sigma,f_0)  = \int_{r_{\rm l}}^{r_{\rm h}} p(r_{\rm obs}|r)p_{\rm Nur,res}(r,t_{\rm \star,obs}|\mu,\sigma,f_0)dr\nonumber \\ & + \int_{-\infty}^{r_{\rm l}} p(r_{\rm obs}|r)p_{\rm Nur,nonres}(r,t_{\rm \star,obs}|\mu,\sigma,f_0)dr + \int_{r_{\rm h}}^\infty p(r_{\rm obs}|r)p_{\rm Nur,nonres}(r,t_{\rm \star,obs}|\mu,\sigma,f_0)dr
  \end{align}
  \begin{align}
  \label{eqn:rch}
  p(r_{\rm obs}, t_{\rm \star,obs}|f) &= \int_{r_{\rm l}}^{r_{\rm h}} p(r_{\rm obs}|r)p_{\rm Ch,res}(r,t_{\rm \star,obs}|f)dr\nonumber \\ & + \int_{-\infty}^{r_{\rm l}} p(r_{\rm obs}|r)p_{\rm Ch,nonres}(r,t_{\rm \star,obs}|f)dr + \int_{r_{\rm h}}^\infty p(r_{\rm obs}|r)p_{\rm Ch,nonres}(r,t_{\rm \star,obs}|f)dr,
\end{align}
where Eqn. \ref{eqn:rnur} is for the Nurture hypothesis and Eqn. \ref{eqn:rch} is for the Chance hypothesis. The probability $p_{\rm Nur,nonres}(r,t_{\rm \star,obs}|\mu,\sigma,f_0)$ includes everything other than $p(R_{\rm obs}|R)$ in the $R=0$ case of Eqn \ref{eqn:resnur}, and $p_{\rm Nur,res}(r,t_{\rm \star,obs}|\mu,\sigma,f_0)$ includes everything other than $p(R_{\rm obs}|R)$ in the $R=1$ case of that same equation.  Similarly, $p_{\rm Ch,nonres}(r,t_{\rm \star,obs}|f)$ includes everything other than $p(R_{\rm obs}|R)$ in the $R=0$ case of Eqn \ref{eqn:resch}, and $p_{\rm Ch,res}(r,t_{\rm \star,obs}|f)$ includes everything other than $p(R_{\rm obs}|R)$ in the $R=1$ case of that same equation.  The term $p(r_{\rm obs}|r)$ accounts for the uncertainty in the period ratio (derived from uncertainties in the measured periods), and we assume $r_{\rm obs}$ is drawn from a Gaussian distribution with mean $r$ and standard deviation equal to the period ratio uncertainty.

Calculating the odds ratios in this way results in a Nurture to Chance ratio of 2.1 for the RV sample, 1.7 for the \textit{Kepler} giant planet pairs, and 1.1 for the \text{Kepler} small planet pairs.  There are only a handful of planets within 3$\sigma$ of the threshold for being near a 2:1 resonance, so it is unsurprising that these results barely differ from what we obtained before.

\end{document}